\documentclass[useAMS,usenatbib]{mn2e}

\usepackage{epsfig} 
\usepackage{multirow} 

\def\gsim{\;\rlap{\lower 2.5pt \hbox{$\sim$}}\raise 1.5pt\hbox{$>$}\;} 
\def\lsim{\;\rlap{\lower 2.5pt \hbox{$\sim$}}\raise 1.5pt\hbox{$<$}\;} 
\def\ie{{\it i.e. }} 
\def\eg{{\it e.g. }} 
\def\dw{\Delta \omega} 
\def\dS{\Delta S} 
\def\dM{\Delta M} 
\def\dz{\Delta z} 
\def\Mres{M_{\rm res}}

\def\Nsym{\phi_{\rm sym}} 
\def\Nasym{\phi_{\rm asym}}

\title[Growing Healthy Merger Trees] {How to Grow a Healthy Merger Tree}

\author[Jun Zhang, Onsi Fakhouri, and Chung-Pei Ma] {Jun Zhang$^{1}$\thanks{E-mail:jzhang@astro.berkeley.edu}, Onsi Fakhouri$^{1}$, Chung-Pei Ma$^{1}$\\
\\
$^{1}$Department of Astronomy, University of California, Berkeley, CA 94720, USA \\
}

\begin{document}

%\date{Accepted . Received ; in original form }    
\pagerange{ 
\pageref{firstpage}-- 
\pageref{lastpage}} \pubyear{2006}

\maketitle

\label{firstpage} 
\begin{abstract}
  We investigate seven Monte Carlo algorithms -- four old and three new --
  for constructing merger histories of dark matter halos using the extended
  Press-Schechter (EPS) formalism based on both the spherical and
  ellipsoidal collapse models. We compare, side-by-side, the algorithms'
  abilities at reproducing the analytic EPS conditional (or progenitor)
  mass function over a broad range of mass and redshift ($z=0$ to
  15). Among the four old algorithms (Lacey \& Cole 1993, Kauffmann \&
  White 1993, Somerville \& Kolatt 1999, Cole et al 2000), we find that
  only KW93 produces a progenitor mass function that is consistent with the
  EPS prediction for {\it all} look-back redshifts.  The origins of the
  discrepancies in the other three algorithms are discussed.  Our three new
  algorithms are designed to generate the correct progenitor mass function
  at each time-step.  We show that this is a necessary and sufficient
  condition for consistency with EPS at any look-back time.  We illustrate
  the differences among the three new algorithms and KW93 by investigating
  two other conditional statistics: the mass function of the $i_{\rm th}$
  most massive progenitors and the mass function for descendants with $N_p$
  progenitors.
  % Results from $N$-body simulations will be needed to constrain these
  % higher-moment statistics.
\end{abstract}

\section{Introduction}

\label{intro}

In the hierarchical structure formation scenario, dark matter halos grow by
accreting and merging with other halos.  Statistically modeling halo
merger histories is important for understanding a diverse spectrum of
astrophysical processes ranging from galaxy formation, the growth of
super-massive black holes, to cosmic reionization.

Numerical simulations aside, the most frequently used theoretical framework
for studying the build up of dark matter halos is the Press-Schechter (PS)
model \citep{PS74}.  This framework is further developed in the so-called
extended Press-Schechter (EPS) model
(\citealt{BCEK91,LC93,MW96,SMT01,ST02}).  For a descendant halo of a given
mass at redshift $z_0$, the EPS model predicts the average mass spectrum of
its progenitors at a higher redshift $z_1$ (the {\it conditional} or {\it
  progenitor} mass function).

The EPS model provides only statistical information about halo mergers and
does not specify how progenitor halos are to be grouped into descendant
halos.  However, it is often useful, particularly in semi-analytic
modeling, to have actual realizations of the merging history for a large
set of haloes. A number of Monte Carlo algorithms have been proposed
  for this purpose (see, \eg,
  \citealt{LC93,KW93,SP97,SL99,SK99,C00,C08,MS07,ND08b}).  These algorithms
  allow one to produce realizations of halo merger trees stretching back to
  high redshifts in a fraction of the time that is required for performing
  and analyzing cosmological $N$-body simulations of comparable
  resolution.

Thus far, most of the commonly used Monte Carlo methods are based on the
spherical EPS theory.  In \citealt{LC93} (also see \citealt{BCEK91}), halo
mergers at each time step are assumed to be binary: one of the progenitor
masses is randomly drawn from the conditional mass function, and the other
progenitor mass is defined by the difference between the descendant halo
mass and this first progenitor mass.  Though this seems to be the most
natural way to generate halo merger histories, it has been pointed out by
several authors that the binary picture does not reproduce the EPS
progenitor abundance at earlier times (see, \eg, \citealt{SK99}).  Moreover,
this problem does not disappear when the time step is greatly reduced. This
fact has led to the investigation of alternative Monte Carlo algorithms
with different recipes for building halo merger trees in the spherical EPS
framework. For example, Somerville \& Kolatt (1999) find that if the binary
assumption is relaxed while taking into account the contribution of mass
from continuous accretion then the progenitor abundance at large look-back
times is better reproduced.  Cole et al. (2000), on the other hand, include
diffuse accretion but preserve the assumption of binary mergers.  More
recently, partially due to the rapid advances in N-body simulation, various
other algorithms have been proposed that are either designed to fit
  N-body results (\eg, \citealt{PCH08,C08,ND08a}) or are based on the
  spherical \citep{ND08b} or ellipsoidal \citep{MS07} excursion set model.
The presence of these numerous Monte Carlo algorithms suggests that building a
Monte Carlo algorithm that is fully consistent with the underlying EPS
model is not unique and can be non-trivial.

We were motivated to write this paper for a number of reasons.  
First, this is a sequel to our previous work \citep{ZMF08}, which
presented an accurate analytic formula for the conditional mass function
for small time-steps in the ellipsoidal EPS model.  This formula is
particularly useful as an input for high-resolution Monte Carlo simulations
of halo merger trees.  Earlier formulae (e.g. \citealt{ST02}) were accurate
only for larger look-back redshifts ($z_1-z_0 \ga 0.1$).  Taking such a
large time-step would limit the dynamic range in both the progenitor mass
and redshift that can be covered in a Monte Carlo simulation.  In addition,
until recently, all previous Monte Carlo algorithms were studied in the
framework of the spherical EPS model, which is well known to produce
inaccurate total (i.e. unconditional) halo mass function when compared with
simulations.  This paper will investigate the algorithms in the 
ellipsoidal model using the formula in \cite{ZMF08}.

Second, as we began to investigate the various Monte Carlo algorithms
proposed in the literature, we were frustrated by the lack of direct
comparison among the different methods, each of which has its own range of
validity and own set of assumptions about how to group progenitors into
descendants (e.g. binary vs multiple progenitors; how the mass in
progenitors below mass resolution is treated).  Moreover, it was not always
clear why a given algorithm succeeded or failed.  In this paper, we examine
closely the four most frequently used algorithms -- \citealt{LC93, KW93,
  SK99, C00} -- and compare their predictions for the conditional mass
function over a wide range of progenitor masses and look-back redshift
(e.g., down to $10^{-6}$ of descendant mass and up to redshift 15, much
larger than those studied previously).  We find that only \cite{KW93} is
fully consistent with the EPS model at all look-back time steps.  The
limitations and causes of discrepancies in the other three methods are
discussed.

Third, in light of the discrepancies in earlier algorithms, we investigate
three new Monte Carlo algorithms that are all constructed to reproduce
accurately the EPS predicted conditional mass function at any look-back
redshift.  We present a consistency criterion that is useful as a general
guide for building Monte Carlo algorithms: If an algorithm reproduces the
EPS progenitor mass function for a sequence of simulation time-steps between
$z_i$ and $z_{i+1}$ (where $i=0,N$), then it is guaranteed to reproduce the
EPS progenitor mass function at any $z_j$ for descendants at any later $z_k$
(where $j,k=0,N$).  This is a necessary and {\it sufficient} condition.

Fourth, the EPS model is an incomplete theory that predicts only a subset
of statistical properties of halo mergers.  It therefore leaves one with
much freedom in how to assign progenitors to descendants in a given Monte
Carlo algorithm. For instance, it is possible to construct different
consistent Monte Carlo algorithms that predict different statistical merger
quantities beyond the conditional mass function.  Our three new algorithms
and KW93 are four examples that are degenerate in the conditional mass
function but are different in other progenitor statistics.  In this paper
we illustrate the differences among the models with two such statistics:
the mass function of the $i^{th}$ most massive progenitors and the mass
function of progenitors for descendant halos with $N_p$ progenitors.
Results from $N$-body simulations will be needed to constrain these
higher-moment statistics. Since computing the statistics of progenitor
  dark matter halos in simulations is by itself a major independent
  project, we will focus on the EPS theory and Monte Carlo algorithms in
  this paper and leave the comparison with $N$-body results to a subsequent
  paper (Zhang, Fakhouri \& Ma 2008, in prep).

The paper is structured as follows.  The EPS formalism based on both the
spherical and ellipsoidal gravitational collapse models is reviewed in
\S\ref{EPS}.  In \S\ref{MCEPS} we discuss three ingredients for how to grow an
accurate Monte Carlo merger tree: the consistency criterion for reproducing
EPS (\S\ref{theorem}), the asymmetry in the EPS progenitor mass function
and the necessity of non-binary mergers in an algorithm (\S3.2), and the
role of mass resolution and diffuse accretion for progenitor mass
assignment (\S3.3).  Details of the four old and three new algorithms are
discussed in \S4 and \S5, respectively.  Whenever possible, the resulting
progenitor mass functions from different algorithms are shown on the same
plots for ease of comparison.  \S6 compares the two new progenitor
statistics that can be used to discriminate among the Monte Carlo
algorithms that are consistent with EPS.  We summarize our findings in
\S\ref{summary}, with a discussion of some recent work in this field.

The calculations in this paper assume a $\Lambda$CDM model with
$\Omega_m=0.25$, $\Omega_b=0.045$, $h=0.73$, $\Omega_{\Lambda}=0.75$,
$n=1$, $\sigma_8=0.9$. This is the same cosmology used in the Millennium
simulation (\citealt{springel05}).

\section{An Overview of EPS} 
\label{EPS}

In this section we present a brief overview of the EPS theory based on both
the spherical and ellipsoidal gravitational collapse models. We often
  refer to the two models in parallel as the spherical and ellipsoidal EPS
  models, with the understanding that the ellipsoidal version is based on
  the excursion set formalism of \cite{BCEK91}.  The emphasis here is on
the conditional mass function, which is the main statistical quantity used
to generate progenitors in merger tree algorithms. For a more complete and
pedagogical review of EPS, see \cite{zentner07} and references
  therein.

\subsection{EPS Based on the Spherical Collapse Model} 
\label{basics}

The Press-Schechter (PS) model provides a framework for identifying
virialized dark matter halos. It is assumed that the seed density
perturbations that grow to form these halos are characterized by an
initially Gaussian random density field with larger fluctuations on smaller
spatial scales. This latter assumption allows one to use
$S(R)=\sigma^2(R)$, the variance of the linear density
fluctuations\footnote{In this paper, the variance of the density
  fluctuation is calculated using the fitting formula of the linear mass
  power spectrum from \citealt{EH98}} smoothed over spatial scale $R$, as a
proxy for the spatial scale $R$. Moreover, since a given spatial scale is
related to a unique mass scale $M(R)$ via the mean density of the universe
$\bar{\rho}$, one can use $R$, $M$, and $S$ interchangeably as measures of
scale.

The density field smoothed over a given scale $M$ is given by
$\delta_M=\rho_M/\bar{\rho}-1$ where $\rho_M$ is the average density within
the smoothing scale $R$. In the EPS model, the linear density field centered
at a given point in the initial Lagrangian space traces out a random
walk (referring to a Markovian process)\footnote{Strictly speaking, this is only true when the smoothing
  window function is a top-hat in Fourier space.} as the smoothing scale is
reduced. Starting from a large smoothing scale, a virialized dark matter
halo is assumed to form at the given spatial coordinate when the linear
$\delta_M$ crosses a critical value for the first time; the mass of the
halo is determined by the smoothing scale at first-crossing. In the
spherical EPS model, the critical over-density is given by the spherical
collapse model and is a constant $\delta_c=1.69$ independent of mass scale.

In the above description, as a result of gravitational instability, the
density field grows with time as a linear function of its initial value,
\ie, $\delta_M(z)=\delta_M(0)D(z)$, where $D(z)$ is the standard
cosmology-dependent linear growth factor satisfying $D(z=0)=1$. In
practice, one usually fixes the value of $\delta_M$ at some reference time
(\eg today: $\delta_M(0)$) and evolves the critical over-density to
identify virialized halos at earlier redshifts. We denote this
time-dependent critical over-density by $\omega(z)=\delta_c/D(z)$. Note
that a lower redshift corresponds to a smaller $\omega(z)$, implying that
larger halos form at later times, in accordance with the hierarchical
structure formation scenario.

Under the assumption of Gaussian statistics, the EPS framework allows one
to compute the first crossing distribution $f(S(M_1),z_1|S(M_0),z_0)$. Of
the set of random walks that begin at $\delta_{M_0}=\omega(z_0)$, the first
crossing distribution is the fraction of these random walks that
\emph{first} cross the critical over-density $\omega(z_1)$ at scale
$S(M_1)$, where $z_1>z_0$ and $S(M_1)>S(M_0)$ (\ie $M_1<M_0$). It can be
shown \citep{LC93} that the first crossing distribution in the spherical
EPS model has the form
\begin{eqnarray}
\label{firstcrossingdistribution} 
  &&f(S(M_1),z_1|S(M_0),z_0)d\dS\\ \nonumber
  &=&\frac{1}{\sqrt{2\pi}}\frac{\dw}{{\dS}^{3/2}}
       \exp \left[-\frac{(\dw)^2}{2\dS}\right]d\dS 
\end{eqnarray}
where $\dw=\omega(z_1)-\omega(z_0)$ and $\dS=S(M_1)-S(M_0)$.

The first crossing distribution can be reinterpreted as the conditional
mass function $P(M_1,z_1|M_0,z_0)$, which is defined to be the mass
  fraction of a descendant halo of mass $M_0$ at redshift $z_0$ that
  originates from a progenitor halo of mass $M_1$ at redshift $z_1$:
\begin{equation}
\label{P} 
       P(M_1,z_1|M_0,z_0) dM_1 = -f(S(M_1),z_1|S(M_0),z_0) d\dS 
\end{equation}
Note, in particular, that $P(M_1,z_1|M_0,z_0)$ is the \emph{mass-weighted}
conditional mass function as it represents the merging history of a unit of
mass. The average number of progenitors of mass $M_1$ at $z_1$ associated
with the formation of \emph{each} descendant halo of mass $M_0$ at $z_0$ is
given by the \emph{number-weighted} conditional mass function
$\phi(M_1,z_1|M_0,z_0)$, which is simply related to the mass-weighted
conditional mass function $P(M_1,z_1|M_0,z_0)$ by
\begin{equation}
\label{dndm} 
   \phi(M_1,z_1|M_0,z_0)\equiv 
   \frac{M_0}{M_1} P(M_1,z_1|M_0,z_0) \,. 
\end{equation}
For brevity, we often refer to the number-weighted conditional mass
function $\phi(M_1,z_1|M_0,z_0)$ as the \emph{progenitor} mass function,
and denote it simply as $\phi(M_1|M_0)$ with $z_0$ and $z_1$ specified
elsewhere in paper. This quantity is sometimes denoted as
$dN(M_1,z_1|M_0,z_0)/dM_1$ in the literature.

\subsection{EPS Based on the Ellipsoidal Collapse Model} 
\label{ellipsoidal_theory}

The original Press-Schechter theory was based on the spherical collapse
model. The unconditional mass function in this model is well known to have
an excess of small halos and a deficit of massive halos in comparison with
simulation results (\eg, \citealt{LC94,GB94,MB94,T98,ST99}). This discrepancy
arises because halo collapses are generally triaxial rather than spherical
(\citealt{d70,bbks86,SMT01,ST02}). In the spherical collapse picture, the
virialization of a dark matter halo is purely determined by the
density-contrast on the scale of the halo mass. This assumption is too
simplistic because dark matter halos generally have non-zero ellipticity
and prolateness, and the condition for virialization should be determined
by both the density-contrast and the halo shape parameters.

By assuming that a dark matter halo virializes when its third axis
collapses, \cite{SMT01} find a new criterion for virialization that depends
on the ellipticity and prolateness of a dark matter halo in addition to its
density contrast. In practice, this condition can be simplified either by
averaging over its dependence on the shape parameters, or by fixing the
shape parameters at their most likely values for a given over-density. By
doing so, these authors obtain a fitting formula for the
scale-\emph{dependent} critical over-density, or barrier, in contrast to
the scale-independent $\delta_c$ of the spherical collapse model. It is
parameterized as (\citealt{ST02}):
\begin{equation}
\label{Barrier}
     \delta_c^E[S(M),z]
=\sqrt{\gamma}\delta_c\left[1+\beta(\gamma\nu)^{-q}\right]
\end{equation} 
where $q=0.615$, $\beta=0.485$, $\gamma=0.75$, $\nu=\omega^2(z)/S(M)$, and
$M$ is the halo mass.  In this ellipsoidal collapse model, the
scale-dependence is such that the formation of small halos is delayed,
thereby reducing their abundance and providing closer agreement with the
unconditional mass function in simulations than the spherical model.

To compute the {\it conditional} mass function in the ellipsoidal EPS
model, one would need the equivalent of the first-crossing distribution
eq.~(\ref{firstcrossingdistribution}). The exact analytical form of
eq.~(\ref{firstcrossingdistribution}), unfortunately, is valid only for the
scale-independent constant barrier $\delta_c$ of the spherical EPS
model. \cite{ST02} have presented a Taylor-series-like approximation for
the ellipsoidal model, but \cite{ZMF08} show that this form works well for
large $z_1-z_0$ but is invalid for small $z_1-z_0$.  As the construction of
an accurate ellipsoidal Monte Carlo merger tree algorithm requires accurate
knowledge of the ellipsoidal progenitor mass function at small time-steps,
it is crucial that this matter be resolved.

This was done in \cite{ZMF08}. Using the scale-dependent critical
over-density of \cite{ST02} and the technique of \cite{ZH06}, \cite{ZMF08}
derived an accurate form for the progenitor mass function of ellipsoidal
EPS model for small time steps ($\dz \lsim 0.1$), which can be written as:
\begin{eqnarray}
\label{dndm_E} 
    \phi(M_1|M_0)&=&\frac{M_0}{M_1}\frac{dS(M_1)}{dM_1}
     \frac{A_0\dw}{\dS\sqrt{2\pi\dS}}\\ \nonumber
      &\times&\left\{\exp\left[-\frac{1}{2\dS} \left(A_0\dw + 
       A_1\sqrt{\dS\tilde{S}}\right)^2\right]\right.\\ \nonumber
      &+&\left.A_2\tilde{S}^{3/2}\exp\left(-\frac{A_1^2}{2}\tilde{S}\right)
      \left[1+\frac{A_1}{\Gamma(3/2)}\sqrt{\tilde{S}}\right]\right\} 
\end{eqnarray}
where $A_0=0.866(1-0.133\nu_0^{-0.615})$, $A_1=0.308\nu_0^{-0.115}$,
$A_2=0.0373\nu_0^{-0.115}$, $\nu_0=\omega^2(z_0)/S(M_0)$, and
$\tilde{S}=\dS/S(M_0)$. Note that unlike eq.~(15) of \cite{ZMF08}, we have
not neglected the small $A_0\dw$ term in the exponent because it is
important for tracing the massive progenitors (small $\dS$).  Two other
features are worth noting. First, unlike in the spherical EPS model,
$\phi(M_1|M_0)$ in eq.~(\ref{dndm_E}) depends weakly on the redshift $z_0$.
Second, due to the intersections of barriers at the low mass end\footnote{see
  appendix A of \citealt{ST02} for more details}, eq.~(\ref{dndm_E}) turns
unphysical (\ie, $A_0<0$) when $S(M_0)\gsim 30\,\omega^2(z_0)$, \ie, when
the descendant mass is much smaller than the typical halo mass at $z_0$. In
our Monte Carlo simulations discussed below, whenever the second feature
becomes a problem (which occurs very rarely), we do not generate any
progenitors for the halo in the next time step.  As we will show in
\S\ref{newmethods}, this procedure only mildly affects the progenitor
abundance at the very low mass end.  Eq.~(\ref{dndm_E}) provides a closer
match to the merger rates determined from $N$-body simulations
\citep{ZMF08}, but the agreement was not perfect, perhaps due to the
non-Markovian nature of numerical simulations.

\section{Ingredients for Growing Healthy Monte Carlo Merger Trees} 
\label{MCEPS}

As discussed in the introduction, the EPS model only provides a subset of
\emph{statistical} information about dark matter halo merger histories. For
example, the EPS progenitor mass function $\phi(M|M_0)$ (eq.~\ref{dndm} for
spherical and eq.~\ref{dndm_E} for ellipsoidal) gives the {\it average}
mass spectrum of the progenitors for the descendant halos. However, it is
often useful, especially in semi-analytical modeling, to have an actual
Monte Carlo realization of the formation history for a large set of
halos. Of particular interest is the merger tree of individual halos, which
provides the hierarchical links among the progenitors and their
descendants. Since the EPS model itself does not specify explicitly how to
group progenitors into descendants, in each time-step in a Monte Carlo
algorithm, assumptions must be made about the number of progenitors and
their mass distributions to be assigned to a given descendant.

The earlier Monte Carlo algorithms (e.g., \citealt{LC93, KW93, SK99,
  C00}) for merger tree constructions share a similar overall structure: A
descendant halo of mass $M_0$ at some redshift $z_0$ (typically $z_0=0$) is
chosen. The EPS progenitor mass function, $\phi(M|M_0)$, is then used to
generate a set of progenitors at some earlier redshift, using the rules of
the given algorithm. In the next time-step, these progenitors become
descendants, and each is assigned its own set of progenitors at an earlier
redshift using $\phi(M|M_0)$. This process is repeated out to some early
redshift and for a (typically large) number of halos of mass $M_0$ at the
starting $z_0$.

The existence of a number of diverse Monte Carlo algorithms (see further
discussion in \S\ref{methods}) in the literature implies that the above
process is, in fact, not unique and can be quite subtle. We now explore
some of these subtleties and the key ingredients for constructing a healthy
merger tree.

\subsection{A Criterion for Consistently Reproducing the EPS Progenitor
  Mass Function}
\label{theorem}

We consider a Monte Carlo algorithm to be {\it consistent} with EPS if the
merger trees it produces can reproduce the EPS progenitor mass function
$\phi(M_1,z_1|M_0,z_0)$ exactly for \emph{any} set of
$\left\{M_1,z_1,M_0,z_0\right\}$ regardless of the number or size of the
simulation time-steps between $z_0$ and $z_1$.

Clearly, to be consistent with EPS, a Monte Carlo algorithm must
\emph{necessarily} reproduce the EPS-predicted $\phi(M|M_0)$ exactly at
\emph{adjacent} time steps. We now show that this is also a
\emph{sufficient} condition for the Monte Carlo method to reproduce
$\phi(M|M_0)$ exactly at \emph{any} look-back time regardless of the
number, or width, of intervening time-steps. This condition is important
because it simplifies the analysis of Monte Carlo algorithms: the failure
of a given algorithm to reproduce faithfully the EPS $\phi(M|M_0)$ at a
particular redshift or mass range necessarily implies that the algorithm
fails to reproduce the progenitor mass function (in either amplitude or
shape or both) across a single time step.

We start with the first crossing distribution
eq.~(\ref{firstcrossingdistribution}) and note that due to the continuous
nature of the random walk, it obeys the following identity at different
look-back times:
\begin{eqnarray}
	\label{ff} &&f(S(M),z|S(M_0),z_0)\\
	\nonumber &=&\int_{S(M_0)}^{S(M)}dS' f(S(M),z|S(M'),z') \, f(S(M'),z'|S(M_0),z_0) 
\end{eqnarray}
for any $z_0<z'<z$. This relationship is true in both spherical and
ellipsoidal EPS models because both variants are based on barrier crossings
of random walks. Note that in the ellipsoidal model, eq.~(\ref{ff}) is a
property of only the exact first-crossing distribution, which is well
represented by eq.~(\ref{dndm_E}) for small look-back times but not the
Taylor-series-like approximation of \cite{ST02}.  Also note that
eq.~(\ref{ff}) may not be strictly satisfied at the very low mass end due
to the intersections of barriers in the ellipsoidal model. As we will show
in \S\ref{newmethods}, this only causes a minor problem on very small mass
scales.

Using eqs.~(\ref{P}) and (\ref{dndm}) to relate $f$ to the progenitor mass
function $\phi$, we then obtain
\begin{eqnarray}
\label{Nf}
  &&\phi(M,z|M_0,z_0)\\ \nonumber
   &=&\int_{M}^{M_0}dM' \phi(M,z|M',z') \, \phi(M',z'|M_0,z_0) \,. 
\end{eqnarray}
Setting $z'=z_0+\dz$ and $z=z_0+2\dz$, we see that eq.~(\ref{Nf}) implies
that if a Monte Carlo method generates progenitors \emph{exactly} according
to the progenitor mass function of EPS at each time step $\dz$, then the
Monte Carlo progenitor mass function should agree with the EPS prediction
at any look-back time $z-z_0$.  We stress that $\phi(M|M_0)$ must be
reproduced exactly, that is, in both the overall shape and normalization of
$\phi(M|M_0)$.  This consistency condition is both necessary and
sufficient.

An additional feature to note is that consistency is possible in the
presence of a mass resolution limit $\Mres$ (discussed further in
\S\ref{massres}). Eq.~(\ref{Nf}) shows that $\phi(M,z|M_0,z_0)$ does not
depend on masses outside of the range $[M, M_0]$. Thus if a Monte Carlo
algorithm reproduces $\phi(M|M_0)$ for all $M>\Mres$ in single time-steps,
it will consistently reproduce $\phi(M|M_0)$ at $M>\Mres$ for any $z-z_0$.

\subsection{The Asymmetry of EPS and Binary Mergers} 
\label{asymmetry}

The simplest way to group progenitors into descendants in a Monte Carlo
algorithm is through binary mergers, \ie, each descendant halo of mass
$M_0$ is composed of two progenitors of mass $M_1$ and $M_0-M_1$. This
assumption is used in, e.g., \cite{LC93}.  This simple scenario, however,
will necessarily fail to reproduce both the spherical and ellipsoidal EPS
progenitor mass functions. This is because if all descendants were the
products of binary mergers, then $\phi(M|M_0)$ would be symmetric about
$M_0/2$ for infinitesimal $\dz$. This is simply not the case in EPS.

\begin{figure}
\centering 
\includegraphics{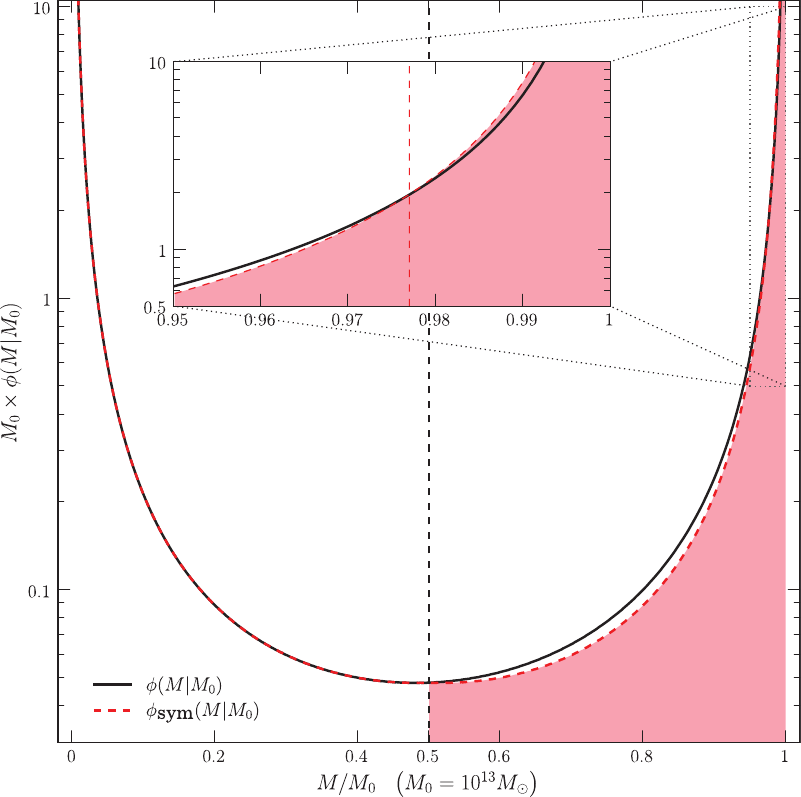}
\caption{An illustration of the asymmetry in the number-weighted
  conditional (or progenitor) mass function $\phi(M,z|M_0,z_0)$ of the
  spherical EPS model for a descendant halo of mass
  $M_0=10^{13}M_{\odot}$ at $z_0=0$ and a look-back redshift of
  $z-z_0=0.02$.  The red dashed curve shows the symmetric part of
  $\phi(M|M_0)$, $\Nsym(M|M_0)$, whose right side is simply the reflection
  of the left side.  The figure indicates that some progenitors of masses
  larger than $M_0/2$ do not have companions in the simplest binary
  scheme. The pop-up is a zoom-in on the right-most part of the plot and
  illustrates that the red dashed curve exceeds the solid curve at
  $M\approx 0.977M_0$.}
\label{spherical_shade}
\end{figure}

We illustrate the asymmetry of the EPS $\phi(M|M_0)$ in
Fig.~\ref{spherical_shade} for a descendant halo of mass $10^{13}M_{\odot}$
 at $z_0=0$ and a look-back time of $z=0.02$ (which is the typical
  time-step used in our Monte Carlo simulations; see Sec.~4).  The solid
black curve shows the total $\phi(M|M_0)$, while the red dashed curve shows
the symmetric part $\Nsym(M|M_0)$ defined by
\begin{equation}
\label{asym} 
	\phi(M|M_0) = \Nsym(M|M_0) + \Nasym(M|M_0) \,, 
\end{equation}
where the left side ($M \le M_0/2$) of $\Nsym(M|M_0)$ is defined to be
identical to $\phi(M|M_0)$ and the right side is defined to be simply the
reflection of the left half about the mid point $M_0/2$. The second term
$\Nasym(M|M_0)$ is then the residual of $\phi$ after subtracting out
$\Nsym$. The figure illustrates that it is not possible for all progenitors
with $M>M_0/2$ to have binary-paired progenitors of mass $M_0-M<M_0/2$. In
particular, we find that for sufficiently small look-back times
(e.g. $z=0.02$ used in Fig.~\ref{spherical_shade}),
$\phi(M|M_0)>\Nsym(M|M_0)$ when $M_0/2 \leq M \lsim 0.97 M_0$ and
$\phi(M|M_0)<\Nsym(M|M_0)$ when $ M \gsim 0.97 M_0$ (see the pop-up in
Fig.~\ref{spherical_shade}). That is, there are slightly fewer progenitors
with masses below $M_0/2$ than above, except near the end points (below
$0.03 M_0$ and above $0.97 M_0$) where the trend is flipped.

Even though the asymmetry is typically small ( $\Nasym \la 0.1\, \Nsym$ out
to $M_0-\Mres$), an accurate algorithm must include non-binary progenitor
events. These can be descendants with either a single progenitor or multiple
($N_p>2$) progenitors, as will be seen in the new algorithms discussed in
\S\ref{newmethods} below. This fact was emphasized by \cite{ND08b}.  These
authors construct a mass conserving consistent Monte Carlo algorithm that
produces a large number of non-binary descendants.  However, one
intuitively expects that more mergers will be binary as $z_1-z_0
\rightarrow 0$.  This intuition is supported by results from the Millennium
simulation \citep{FM08}, which show that the binary assumption becomes
increasingly valid down to smaller $\Mres$ as $z_1-z_0$ is made smaller. 
This result suggests that the Markovian nature of the standard EPS
model with a top-hat smoothing window may need to be modified 
to account for the correlated sequences of mergers occurring in simulations
\citep{ND08b,zentner07}.

\subsection{Mass Resolution, Diffuse Accretion, and Mass Conservation in
  Monte Carlo Algorithms}
\label{massres}

In the EPS model, all the mass in the universe is assumed to be in dark
matter halos\footnote{This is not exactly true in the ellipsoidal EPS
  model. See appendix A of \citealt{ST02}.}. Although the mass-integral of
the (unconditional) mass function in this model is finite, the
number-integral is unbounded; that is, EPS predicts a preponderance of very
low mass halos. Thus, any practical Monte Carlo algorithm must necessarily
assume a lower mass cutoff, the mass resolution $\Mres$.

For a nonzero $\Mres$, a halo's merger history at each time step can be
thought of consisting of mass in the form of resolvable progenitor halos
and a reservoir of mass due to ``diffuse'' accretion that is the aggregate
contribution from all sub-resolution progenitors. This technical
  distinction is introduced for ease of implementing the Monte Carlo
  methods.  It will, however, play a more physical role when we compare the
  results with $N$-body simulations, which has its own mass resolution as
  well as a possibly physical diffuse component consisting of tidally stripped dark
  matter particles.  In this paper we use $\dM$ to denote this diffuse
accretion component, which we define to be
\begin{equation}
	\dM=M_0-\sum_i M_i \,, 
\end{equation}
where $M_i$ are the masses of the progenitors above $\Mres$ and $M_0$ is
the mass of the descendant.

We call a Monte Carlo algorithm \emph{mass conserving} if each descendant
and its progenitors produced by the algorithm satisfies $\sum_i M_i \le
M_0$. Monte Carlo algorithms are generally expected to be mass conserving,
but we note that this is not a necessary condition for reproducing the EPS
progenitor mass function because the latter is a statistical measure of
merger properties. In two of our new algorithms below (methods A and B in
\S\ref{newmethods}), a small fraction of the descendants can have $\sum_i
M_i > M_0$. We allow this to simplify the description and implementation of
our algorithms. We have experimented with redistributing these excess
  progenitors among other descendant halos in a mass-conserving manner and
  found it not to modify significantly the resulting merger statistics.  In
  addition, it may appear that $\sum_i M_i > M_0$ is unphysical.  We have
  found, however, that a non-negligible fraction of halos in $N$-body
  simulations in fact have $\dM < 0$, perhaps as a result of tidal
  stripping.  This point will be discussed in greater detail in our next
  paper.

We note that for a Monte Carlo algorithm that is consistent with EPS, the
mean value of $\dM$ per descendant halo of mass $M_0$ (i.e., averaged over
all descendants in a given time-step) is, by construction, related to the
mass resolution by
\begin{equation}
	\label{masscons} \langle\dM\rangle =\int_0^{\Mres} M \phi(M|M_0) dM \,, 
\end{equation}
For a given $\phi(M|M_0)$ and $\Mres$, $\langle\dM\rangle$ is therefore
specified. The {\it distribution} of $\dM$, however, can differ greatly
among different algorithms; that is, there is much freedom in how to assign
the amount of diffuse accretion to individual descendants in a given
time-step. For instance, \cite{C00} assumes a delta-function distribution
for $\dM$ (see \S\ref{sec:C00} for details), while most of other methods,
including our new methods discussed in \S\ref{newmethods}, have broader
distributions.

\section{Comparison of Four Previous Monte Carlo Algorithms} 
\label{methods}

In this section we examine four existing Monte Carlo algorithms for
generating merger trees: \cite{LC93} [LC93], \cite{KW93} [KW93],
\cite{SK99} [SK99], and \cite{C00} [C00]. This set is by no means complete,
but these are four of the most frequently used algorithms in the
literature.  The purpose here is to compare these well known algorithms
side-by-side and to illustrate the mass and redshift ranges for which each
method succeeds and fails in matching the spherical EPS model.  This 
not only benefits the current users of the methods, but also prepares us for 
incorporating the ellipsoidal EPS model into the successful method (KW93), 
which will be compared with our new methods in Sec.~5.

\begin{figure*}
\centering 
\includegraphics{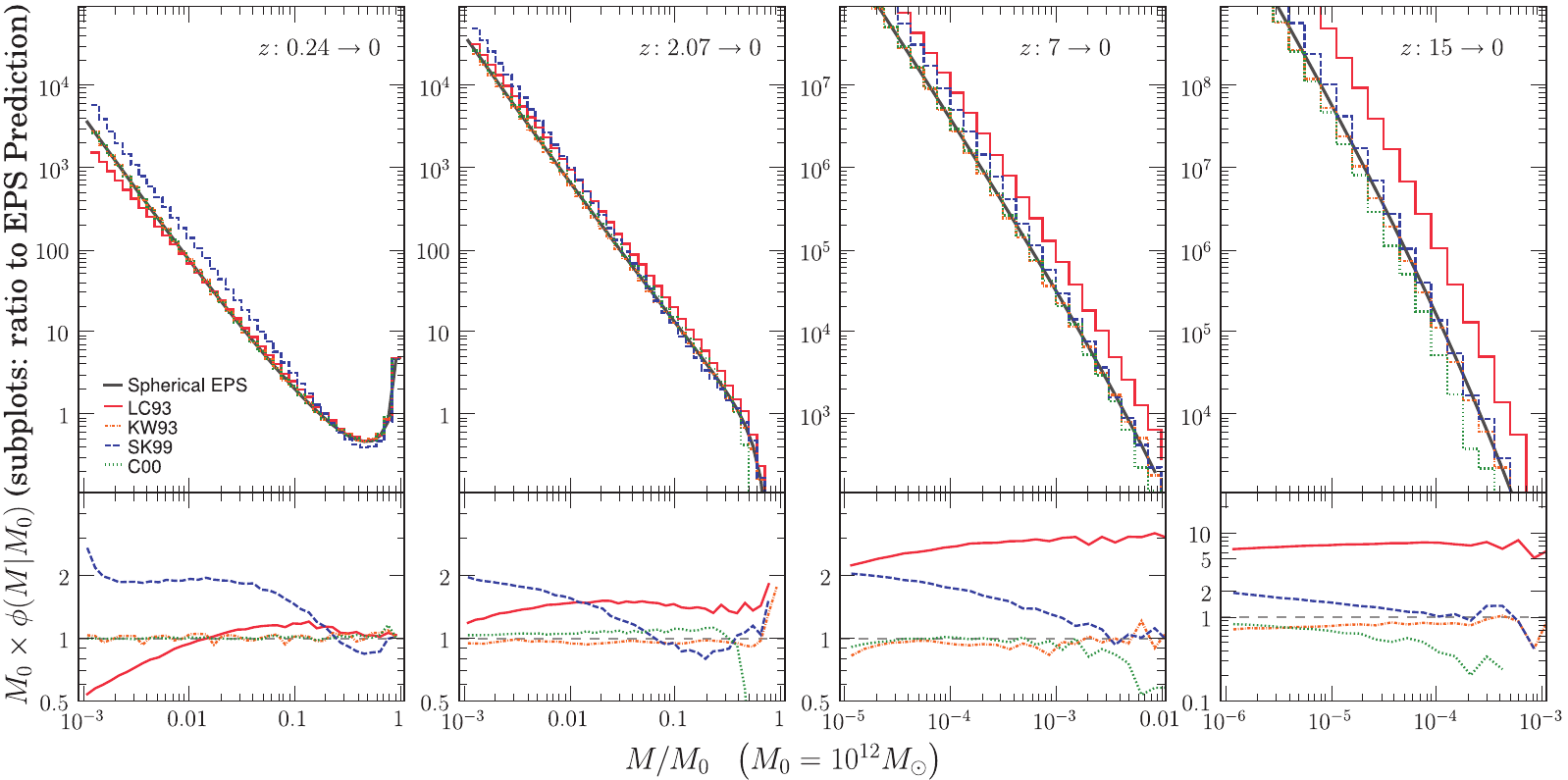} 
\caption{Comparison of the progenitor (or conditional) mass functions
  $\phi(M,z|M_0,z_0)$ that we generated using the four previous Monte Carlo
  algorithms by LC93 (red solid), KW93 (orange dot-dashed), SK99 (blue
  dashed), and C00 (green dotted), and the predictions of the analytic
  spherical EPS model (black solid).  The four panels show four look-back
  redshifts ($z-z_0=0.24, 2.07, 7$ and 15) for a descendant halo of
  $M_0=10^{12} M_\odot$ at $z_0=0$.  For clarity, we plot in the sub-panel
  below each panel the ratios of the Monte Carlo result and the EPS
  prediction.  One can see that KW93 is the only accurate algorithm for all
  $z$.  Note that different ranges of $M/M_0$ are shown in each panel
    since the progenitors have progressively smaller masses at higher
    redshifts.  }
\label{PDF_4EPS_e2}
\end{figure*}

\begin{figure*}
\centering 
\includegraphics{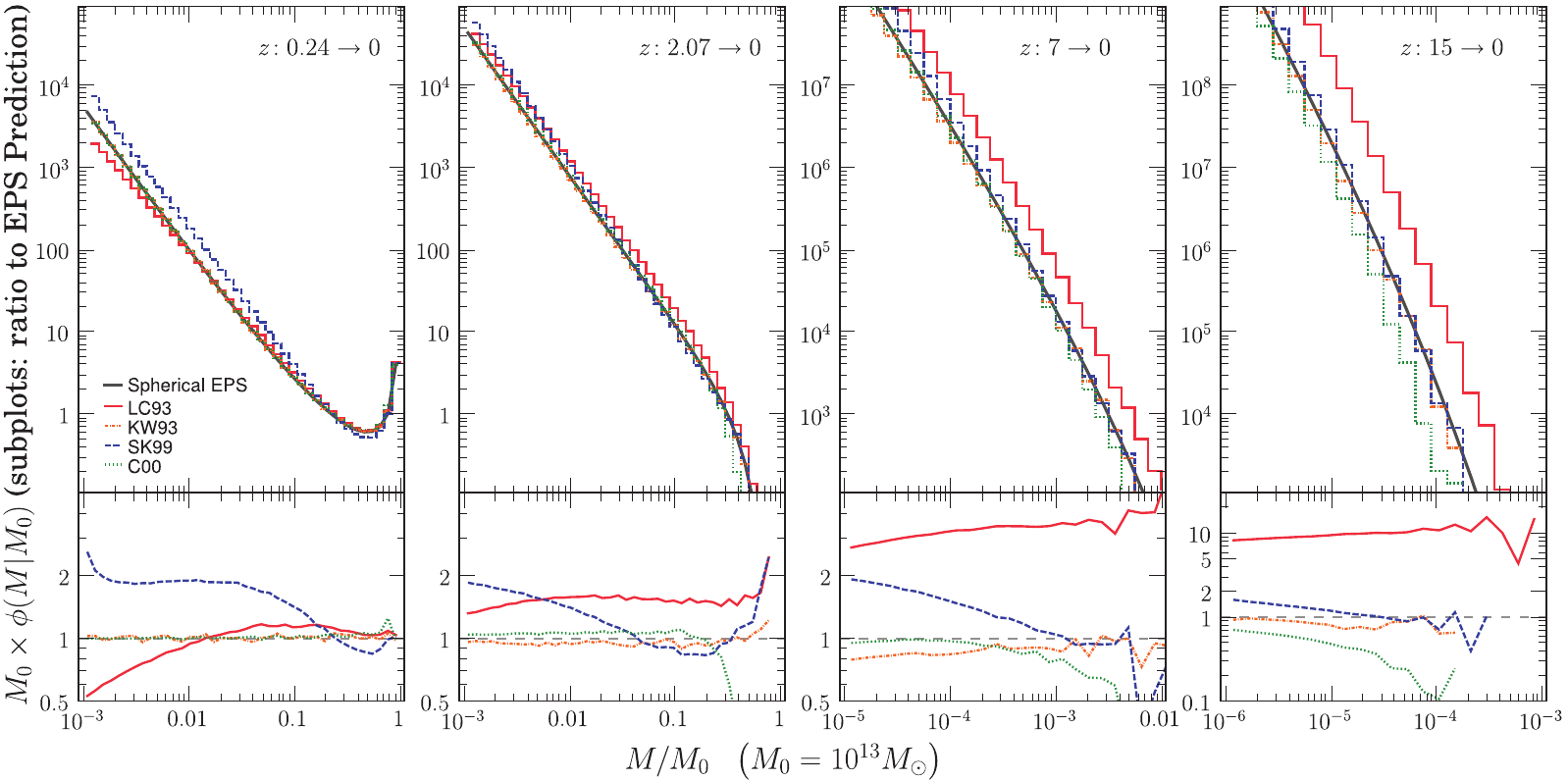} 
\caption{Same as Fig.~\ref{PDF_4EPS_e2}, but for a descendant halo of
  $10^{13}{\rm M}_{\odot}$.} 
\label{PDF_4EPS_e3}
\end{figure*}

\begin{figure*}
\centering 
\includegraphics{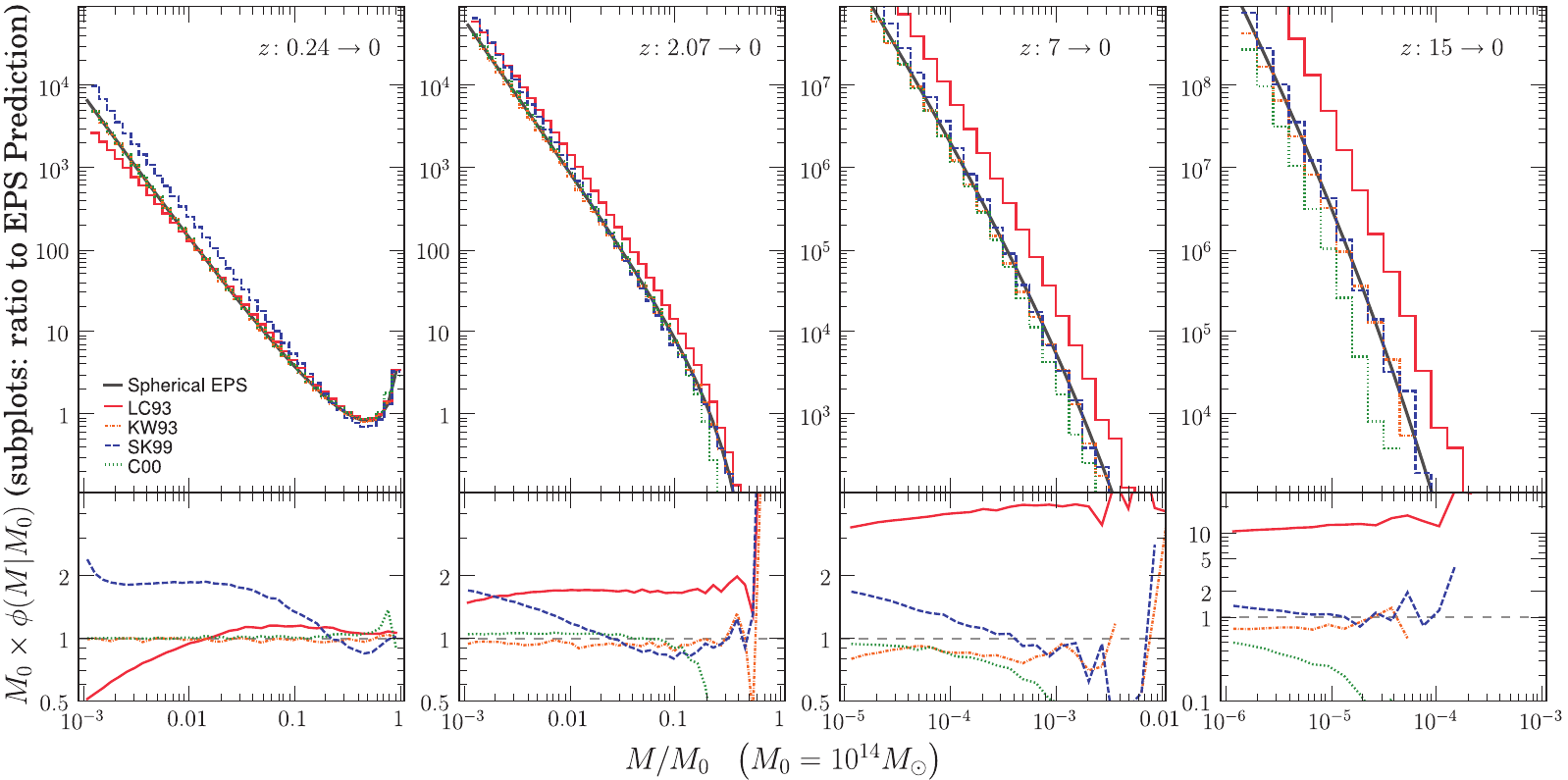} 
\caption{Same as Fig.~\ref{PDF_4EPS_e2}, but for a descendant halo of
  $10^{14}{\rm M}_{\odot}$.}
\label{PDF_4EPS_e4}
\end{figure*}

We review each algorithm in a subsection below and compare the resulting
progenitor mass functions $\phi(M|M_0)$ with the spherical EPS prediction
for look-back redshifts ranging from 0.24 to 15. In
Figs.~\ref{PDF_4EPS_e2}-\ref{PDF_4EPS_e4} we plot the progenitor mass
functions produced by all four methods, along with the analytical EPS
prediction, on log-log plots for three descendant masses ($10^{12},
10^{13}, 10^{14} M_\odot$) and four look-back times ($z_1-z_0=0.24, 2.07,
7$ and $15$). To ease comparison, we also plot the ratio between each Monte
Carlo result and the EPS prediction on a linear-log plot. As
Figs.~\ref{PDF_4EPS_e2}-\ref{PDF_4EPS_e4} clearly show, of the four
algorithms, only KW93 is able to match the spherical EPS $\phi(M|M_0)$ for
{\it all} $z-z_0$.  We will explore why each algorithm fails below and
discuss the care that must be taken when implementing KW93.  A summary of
the four algorithms, their discrepancies, and the causes of the
discrepancies is given in Table~\ref{methodsTableSummary}.

In our Monte Carlo simulations, we generally keep track of all progenitors
down to $0.001 M_0$ at each time step for a descendant halo of mass
$M_0$.  This large dynamic range allows us to predict reliably the
progenitor abundance even for a very large look-back time ($z_1-z_0\sim
15$).  To speed up the algorithm, we take each time step to be a constant
difference in the barrier height $\Delta\omega(z)=\omega(z+\Delta
z)-\omega(z)$ (where $\Delta\omega\approx \Delta z$ at low $z$), which is
chosen to be about $0.02$ for LC93, KW93, SK99, and $0.003$ for C00 at
$z=0$. The progenitor mass function of a given descendant halo mass is then
identical for each time step and does not have to be recomputed. Numerical 
convergence is tested by changing the time-steps
used in the simulation: our results do not change.
\begin{table*}
	\centering 
	\begin{tabular}
		{r|p{1in}|p{2.7in}|p{2.1in}} Algorithm & Overview & Discrepancy in progenitor mass function $\phi(M|M_0)$ & Reasons for Discrepancies \\
		\hline \hline \multirow{2}{*}{LC93} & Binary and 1-to-1 & \multirow{2}{2.7in}{Overestimates $\phi(M|M_0)$ by large factors when the look-back time is large, \ie, $z_1-z_0 \gsim 1$} & \multirow{2}{2.2in}{Binary assumption fails to reproduce EPS $\phi(M|M_0)$ asymmetry.}\\
		&$\dM\leq\Mres$& \\
		\hline \multirow{2}{*}{KW93} & Multiple mergers & \multirow{2}{2.7in}{None} & \multirow{2}{2.2in}{}\\
		&$\dM\neq0$&\\
		\hline \multirow{3}{*}{SK99} & Multiple mergers & \multirow{3}{2.7in}{Typically over-predicts the abundances of small progenitors ($\lsim 10\%$ of the descendant halo mass) by a factor of $\sim 2$ for $z_1-z_0 \lsim 1$. This discrepancy propagates to smaller mass scales for larger look-back times.} & \multirow{3}{2.2in}{Truncation of $\phi(M|M_0)$ fails to reproduce its shape exactly.} \\
		& $\dM\neq0$ (can be bigger or smaller than $\Mres$)& \\
		& & \\
		\hline \multirow{2}{*}{C00} & Binary and 1-to-1 & \multirow{2}{2.7in}{Works reasonably well for a large range of the look-back time but significantly underestimates $\phi(M|M_0)$ at high mass ends, particularly when the look back time is large ($z_1-z_0 \gg 1$).} & \multirow{2}{2.2in}{Binary assumption fails to capture asymmetry of EPS $\phi(M|M_0)$; fixed $\dM$ yields 1-to-1 events that do not accurately reproduce the high mass end of $\phi(M|M_0)$.} \\
		& $\dM$ is a constant given by equation (\ref{C00dM}).& \\
		\hline 
	\end{tabular}
	\caption{A scorecard for the four old Monte Carlo algorithms discussed in \S\ref{methods}. We note that the 1-to-1 events in LC93 and C00 are actually binary mergers involving a secondary progenitor with mass below $\Mres$. Since these progenitors are below the resolution limit they are not counted as progenitors but as diffuse mass $\dM$.} \label{methodsTableSummary} 
\end{table*}

\subsection{Lacey \& Cole (1993)}

The algorithm proposed by LC93 makes two important assumptions: all mergers
are binary (before mass resolution is imposed), and the descendant mass
$M_0$ is the sum of the two progenitor masses $M_1$ and $M_2$ (where $M_1
\ge M_2$ in our convention). For each small look-back time step and for
each descendant, a progenitor mass is randomly chosen according to the
mass-weighted conditional mass function eq.~(\ref{P}), and the mass of the
other progenitor (which can be larger or smaller) is simply set to be the
difference between $M_0$ and the first chosen progenitor mass. If the less
massive progenitor $M_2$ falls below a chosen mass resolution $\Mres$, or
equivalently, $M_1 > M_0 - \Mres$, then $M_1$ is kept but $M_2$, being a
sub-resolution progenitor, is discarded. This results in single-progenitor
halos which we label as ``$1\rightarrow1$'' events. In this notation,
binary mergers are ``$2\rightarrow1$'' events. When a smaller time-step is
used in LC93, the ratio of $2\rightarrow1$ to $1\rightarrow1$ events
decreases.

We find that random progenitor masses can be easily generated using the
parameter transformation:
\begin{equation}
\label{define_x} 
   x={\rm erf}\left\{\dw/\sqrt{2[S(M_1)-S(M_0)]}\right\} \,. 
\end{equation}
The parameter $x$ has a uniform probability distribution between 0 and 1
and can be quickly generated using any random-number generator. A simple
inversion then yields progenitors distributed according to the
mass-weighted conditional mass function.

The red solid histograms and curves in Fig.~\ref{PDF_4EPS_e2} --
\ref{PDF_4EPS_e4} compare the progenitor mass functions generated using the
LC93 algorithm with the predictions of the spherical EPS model (solid black
curves). For all three descendant halo masses shown ($10^{12},10^{13}$ and
$10^{14} M_\odot$), we see close agreement for small look-back times such
as $z_1-z_0=0.24$, but LC93 produces an excess of progenitors at larger
look-back times, and the discrepancy worsens, reaching an order of
magnitude by $z_1-z_0=15$. We believe this discrepancy is due to the binary
nature of LC93: the number of progenitors with mass $M$ is equal to the
number of binary companions of mass $M_0-M$. Thus the LC93 Monte Carlo
algorithm generates a progenitor mass function after one time step that is
symmetric in the left and right sides, which will not match the asymmetric
nature of the EPS $\phi(M|M_0)$ discussed in Sec.~\ref{asymmetry} and shown
in Fig.~\ref{spherical_shade}.  This discrepancy is amplified after many
time-steps when the look-back time becomes large.

Finally, we note that the authors of LC93 also consider another way of
drawing the first progenitor mass from the mass-weighted conditional mass
function, which is to draw it from the mass range of $[M_0/2, M_0]$ instead
of $[0, M_0]$. In practice, we find that this slightly modified version of
LC93 generates very similar results, and our above discussion is valid.

\subsection{Kauffmann \& White (1993)} 
\label{KW93sec}

For each time-step in the KW93 algorithm, a large number of progenitors are
generated across many progenitor mass bins for a fixed number of descendant
halos of the same mass. The number of progenitors in each mass bin is
determined by the progenitor mass function of the descendant halo mass, and
rounded to the nearest integer value. These progenitors are then assigned
to the descendant halos in order of decreasing progenitor mass. The target
descendant halo is chosen with a probability proportional to its available
mass (i.e. the mass not yet occupied by progenitors), and with the
restriction that the total mass of the progenitors in a descendant halo
cannot exceed the descendant mass. This procedure allows one to work out
all the merger configurations and their frequencies for one time step and
for different descendant halo masses. This information is then stored and
used repeatedly for determining the progenitors of a halo at each time
step.

To speed up the implementation of KW93, we divide the look-back time into
steps with equal spacing in the barrier height $\Delta\omega$ as discussed
earlier. The progenitor mass function for a fixed descendant halo mass is
then identical for every time step and only has to be calculated once. We
store the ensemble of progenitors and their merger configurations for each
descendant halo mass bin.  In a Monte Carlo simulation, we randomly select
one merger configuration from the many stored ones for a descendant halo at
each time step.

In practice, we find that extreme care must be taken to avoid numerical
problems in KW93. First of all, this algorithm requires a large number of
progenitor mass bins in the neighborhood of $M_0$ because $\phi(M|M_0)$ is
sharply peaked near $M_1\sim M_0$ for small time-steps. Interestingly, we
find that if the mass range of $[\Mres, M_0]$ is simply divided into
evenly-spaced logarithmic bins, this method is not accurate even when the
number of mass bins is as large as 2000, which already requires more than
$\sim 50000$ descendant halos to guarantee that the integer rounding does
not introduce a significant error to the progenitor number in each bin. As
a result, a large amount of computer memory is necessary to repeat this
procedure for descendant halos of different masses. The improved mass bin
configuration that we end up using will be introduced in
\S\ref{newmethods}. Using that setup, we find that only 200 bins are
required to reproduce accurately the EPS progenitor mass function over
large look-back times.

The second problem is that KW93's scheme for assigning progenitors to
descendant halos is somewhat ambiguous and does not guarantee that all the
progenitors can be assigned. Fortunately, we find that this problem usually
does not arise when the ensemble of progenitors is large. For each
descendant halo mass, we use $\sim 8000$ descendant halos to determine the
merger configurations of the progenitors.

The orange dash-dotted curves in Fig.~\ref{PDF_4EPS_e2} - \ref{PDF_4EPS_e4}
compare the progenitor mass functions generated using the KW93 algorithm
with the predictions of the spherical EPS model (black). The results show
very good agreement. Since KW93 reproduces the exact EPS progenitor mass
function at every time-step, it is expected to be consistent with EPS at any
$z_1-z_0$ according to the discussion in \S\ref{theorem}.

\subsection{Somerville \& Kolatt (1999)} 
\label{sec:sk99}

Somerville \& Kolatt (1999) [SK99] point out that the assumptions of binary
mergers and $M_0=M_1+M_2$ made in LC93 lead to an overestimate of the
progenitor abundance at high redshift. They first attempt to remedy this
problem by preserving the binary assumption while allowing the mass below
the resolution limit $\Mres$ to be counted as diffusely accreted mass $\dM$
(see \S\ref{massres}). They show, however, that this ``binary tree with
accretion'' method fails in the opposite direction, {\it under-producing}
the progenitor mass function relative to the spherical EPS prediction. This
discrepancy arises partly because whenever two progenitors are chosen in
this method, the remaining mass is assigned to $\dM$ regardless of whether
it is above or below $\Mres$. Thus the EPS $\phi(M|M_0)$ is not faithfully
reproduced: the binary tree with accretion method yields an excess of
accreted mass and a corresponding shortage of low-mass halos.

SK99 then consider a natural extension of this method, in which both
assumptions made in LC93 are relaxed.  In this ``N-branch tree with
accretion'' algorithm, each descendant halo is allowed to have more than
two progenitors for every simulation time-step. To guarantee that the total
mass of the progenitors does not exceed that of the descendant, each
subsequent progenitor mass is randomly chosen from the mass-weighted
conditional mass function truncated to the maximally possible progenitor
mass. This procedure is repeated until the descendant halo cannot contain
any more progenitors with masses above $\Mres$, and the remaining mass
deficit is assigned to diffuse accretion $\dM$.

The parameter transformation of eq.~(\ref{define_x}) is also applicable for
SK99. The probability distribution of $x$ is still uniform, but the upper
limit of $x$ can now take on any value between 0 and 1 depending on where
the conditional mass function is truncated.

The blue dashed curves in Fig.~\ref{PDF_4EPS_e2} - \ref{PDF_4EPS_e4}
compare the progenitor mass functions generated using the N-branch tree
algorithm of SK99 with the predictions of the spherical EPS model
(black). It is interesting to note that the sign of the discrepancy is now
opposite to that of LC93: SK99 produces an excess of low-mass ($\lsim 0.1
M_0$) progenitors by up to a factor of $\sim 2$ for small look-back times,
but it does a better job than LC93 at high redshifts. However, it is
noteworthy that even at high redshifts, discrepancies of up to a factor of
$\sim 2$ are still present for small progenitor masses.

We believe that the use of a truncated progenitor mass function in SK99 is
at least a partial cause for the over-prediction of small
progenitors. Since the distribution of progenitors (in particular, the
upper limit for the progenitor mass) depends on the sum of the masses of
the progenitors already picked out for the current halo, the \emph{order}
in which progenitor halos are randomly pulled out matters in this
method. Halos more massive than the truncation limit are effectively
discarded instead of being randomly selected and placed in, for example,
new descendant halos. This procedure tends to preferentially skew the
progenitor mass function at small time steps towards more low mass
progenitors and fewer high mass progenitors.

\subsection{Cole et al. (2000)} 
\label{sec:C00}

Similar to SK99, \cite{C00} [C00] treats the mass in progenitors smaller
than the mass resolution $\Mres$ in the Monte Carlo simulation as accreted
mass, but unlike the N-branch tree model in SK99, only a maximum of two
progenitors are allowed per descendant. The amount of accreted mass gained
in one time-step, $\Delta M$, is fixed to a single value and is calculated
by integrating the mass-weighted conditional mass function from 0 to
$\Mres$:
\begin{equation}
\label{C00dM} 
   \Delta M = \int_0^{\Mres} M \phi(M|M_0) dM \,, 
\end{equation}
where $M_0$ is the descendant mass. The progenitors are drawn from the {\it
  lower} half of the progenitor mass function between $\Mres$ and $M_0/2$
according to the average number of progenitors in that range:
\begin{equation}
	p = \int_{\Mres}^{M_0/2} \phi(M|M_0) dM \,. 
\end{equation}
The simulation time-step is chosen to be small enough so that $p\ll 1$ (note
that it is for this reason that we use $\dz=0.003$ when implementing C00). 

The C00 merger tree is generated with the following steps: A random number
$x$ between 0 and 1 determines whether a descendant halo has one progenitor
(if $x > p$) or two progenitors (if $x \leq p$). In the case of a single
progenitor, its mass is $M_1=M_0-\Delta M$. In the case of two progenitors,
the mass of the smaller progenitor, $M_2$, is chosen randomly between
$\Mres$ and $M_0/2$ according to the progenitor mass function. The larger
progenitor is then assigned a mass of $M_1=M_0-M_2-\Delta M$. Since $p\ll
1$, most descendants form via $1\rightarrow1$ events rather than
$2\rightarrow1$ events. To improve the speed of this algorithm, we precompute 
and store the binary merger rates and diffuse accretion mass fractions for 
a single time step for different descendant mass bins. 

The green dotted curves in Fig.~\ref{PDF_4EPS_e2} - \ref{PDF_4EPS_e4}
compare the progenitor mass functions generated using the C00 algorithm
with the predictions of the spherical EPS model (black). The agreement is
noticeably better than LC93 and SK99. The largest discrepancy occurs at the
high mass end at large $z_1-z_0$, where C00 {\it under-predicts} the
progenitor number at $z_1$ by more than a factor of two for
group-to-cluster size descendants at $z_0$ with $M_0 \ga 10^{13} M_\odot$.

At least two problems contribute to this discrepancy: (i) Since $\dM$ is
fixed to one value (eq.~\ref{C00dM}), the mass of the progenitor for
$1\rightarrow1$ descendants is also a fixed value: $M_1=M_0-\Delta M$. This
is an over-simplification that compresses the high mass end of
$\phi(M|M_0)$ into a delta function. (ii) For descendants with binary
progenitors, C00 uses the spherical EPS conditional mass function only in
the lower mass range $[0, M_0/2]$ to generate the progenitor abundance. By
construction, then, the shape of the progenitor mass function in the upper
mass range, $[M_0/2, M_0]$, is symmetric with the lower half and fails to
match accurately the asymmetric EPS $\phi(M|M_0)$.

\section{Three Consistent Monte Carlo Algorithms} 
\label{newmethods}

In this section, we present three Monte Carlo algorithms that all satisfy
the criterion for consistency discussed in \S\ref{theorem} and will
therefore accurately reproduce the EPS progenitor mass function
$\phi(M|M_0)$. We introduce the common setup for our methods in
\S\ref{commonsetup} and discuss in detail how each method assigns the
ensemble of progenitors to descendants in \S\ref{methodA} -- \ref{methodC}.

To help the reader follow our discussions, we provide a summary of the
breakdown of the merger configurations for the three new algorithms in
Table~\ref{newMethodsTableSummary} and the accompanying
Fig.~\ref{fig:methodsExplanation}.

Although the standard practice in the community has been to generate merger
trees using the spherical EPS model, we emphasize that the Monte Carlo
algorithms can be applied to the ellipsoidal EPS model as well. In fact,
since the ellipsoidal model matches the unconditional mass function in
simulations better than the spherical model, we would expect the
ellipsoidal EPS to also match better the progenitor statistics in
simulations. We will therefore present our results for both the spherical
and ellipsoidal EPS models in parallel below.

\subsection{The Common Setup} 
\label{commonsetup}

\subsubsection{Basic Features}

Our Monte Carlo algorithms for growing consistent merger trees all share
the following implementation framework. We begin at redshift 0 and build
the merger tree backwards in cosmic time. We typically choose a large
descendant halo mass range ($M_0 = [10^{6}M_{\odot}, 10^{15}M_{\odot}]$)
and a small simulation time-step ($\Delta z\approx 0.02$ at low $z$; see
discussion below) to achieve a high resolution tree and a large dynamic
range in the progenitor mass. For a given descendant halo, we first compute
which mass bin it belongs to, and then obtain its progenitors across a
single time-step using the distribution of merger configurations specific to
each algorithm (described in the next three subsections). The progenitors
then become descendants in the next time-step, and this process is repeated
to build up the higher tree branches.

To be specific, a merger configuration here is defined as a set of
progenitor masses that form a descendant halo of a given mass in one
time-step. For example, for a descendant halo of mass $M_0$, one merger
configurations may include only two progenitors of mass $0.6M_0$ and
$0.4M_0$, while another may contain three progenitors of mass $0.4M_0$,
$0.3M_0$, and $0.2M_0$. Note that the sum of the progenitor mass in each
configuration need not equal the descendant mass, and the deficit, $\dM$,
is implicitly attributed to sub-resolution progenitors (see
\S\ref{massres}). Different Monte Carlo algorithms have different
distributions of merger configurations and progenitor multiplicities for
each descendant bin. For convenience, we call the most massive progenitor
in a merger configuration the {\it primary} progenitor, and the rest of the
progenitors the {\it secondary} progenitors.

Our basic implementation is applicable to both the spherical and
ellipsoidal EPS models. We find a particularly efficient choice of time-step
to be the one corresponding to a constant difference in the barrier height
$\Delta\omega(z)=\omega(z+\Delta z)-\omega(z)$, as is used in
\S\ref{methods} for the four old algorithms. For the spherical case, the
progenitor mass function eq.~(\ref{dndm}) depends on time only through
$\Delta\omega(z)$ and is therefore identical for all redshifts when the
same $\Delta\omega(z)$ is used. Thus we only have to generate the merger
configurations in the spherical case across a single time-step once. For the
ellipsoidal case, however, the progenitor mass function
eq.~(\ref{dndm_E}) not only is a function of $\Delta\omega(z)$ but also
depends explicitly on $z$. For each Monte Carlo algorithm, it is therefore
necessary to generate and store the merger configurations and their
probabilities for both descendant halos of different masses and several
redshift bins. In practice, since the redshift dependence of
eq.~(\ref{dndm_E}) is weak, typically fewer than $\sim20$ redshift bins are
required.

\subsubsection{Important Progenitor Mass Scales} 
\label{mass_scales}

A number of natural mass boundaries play critical roles in the construction
of our algorithms. These mass scales demarcate the regions with different
progenitor multiplicities, as illustrated in
Fig.~\ref{fig:methodsExplanation} and discussed in detail in the next three
subsections.

(i) The resolution scale $\Mres$ and its complement $M_0-\Mres$ are two
obvious boundaries, as is the half descendant mass $M_0/2$ discussed in the
context of binary mergers in Sec.~\ref{asymmetry}. We generally choose a
small $\Mres$ (typically $\Mres=0.001 M_0$) for numerical precision and
keep track of all the progenitors down to this limit at each time-step.

(ii) The mass $\alpha M_0$ given by 
\begin{equation}
\label{alpha}
	\int_{\alpha M_0}^{M_0} \phi(M|M_0)\,dM=1 \,
\end{equation}
defines the range of progenitor mass over which every descendant halo is
guaranteed to have one progenitor with $M \in [\alpha M_0, M_0]$.
Table~\ref{boundaryStats} lists the values of $\alpha$ for both the
spherical and ellipsoidal progenitor mass functions for three descendant
masses; $\alpha$ is seen to range from 0.361 to 0.448.

(iii) The mass $\mu M_0$ demarcates where the asymmetric progenitor mass
function self-intersects: $\phi(\mu M_0|M_0)=\phi(M_0-\mu M_0|M_0)$ with
$\mu>0.5$. For binary merger configurations of the form $M_0=M_1+M_2$,
$\phi(M_1|M_0)>\phi(M_2|M_0)$ when $M_1<\mu M_0$ and
$\phi(M_2|M_0)>\phi(M_1|M_0)$ when $M_1>\mu M_0$.  This mass scale is
illustrated in the pop-up in Fig.~\ref{spherical_shade}.
Table~\ref{boundaryStats} shows that $\mu \approx 0.956$ to 0.977.

\begin{figure}
\centering 
\includegraphics{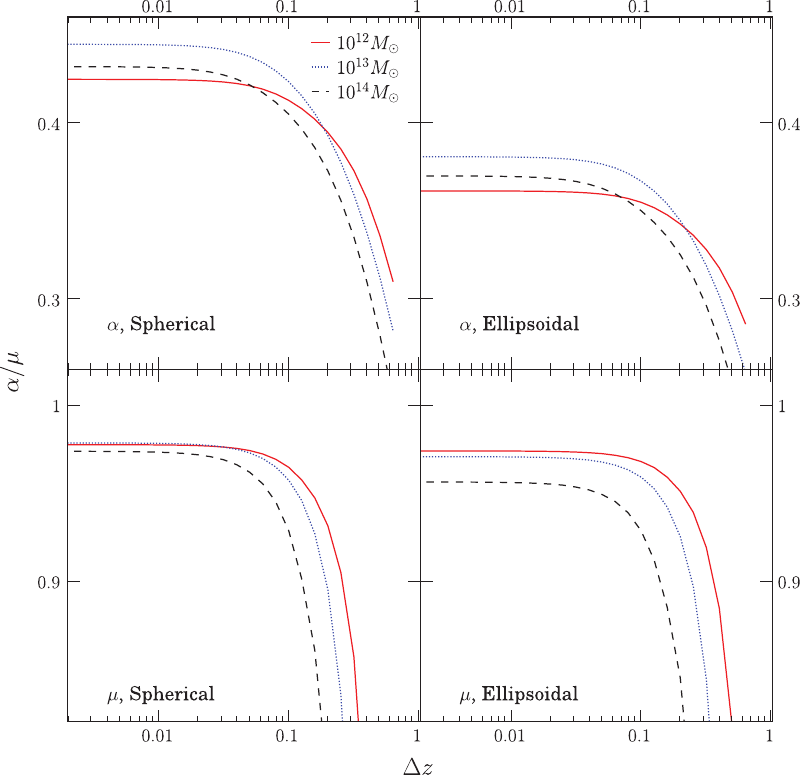} 
\caption{$\alpha$ and $\mu$ as functions of the look-back time $\Delta z$ 
at redshift zero. The red solid, blue dotted, and black dashed curves are for 
descendant halos of $10^{12}M_{\odot}$, $10^{13}M_{\odot}$, and $10^{14}M_{\odot}$ 
respectively. The label in each plot indicates the quantity ($\alpha$ or $\mu$) 
shown and the EPS model (spherical or ellipsoidal) used.
}
\label{alpha_mu}
\end{figure}

\begin{figure}
\centering 
\includegraphics{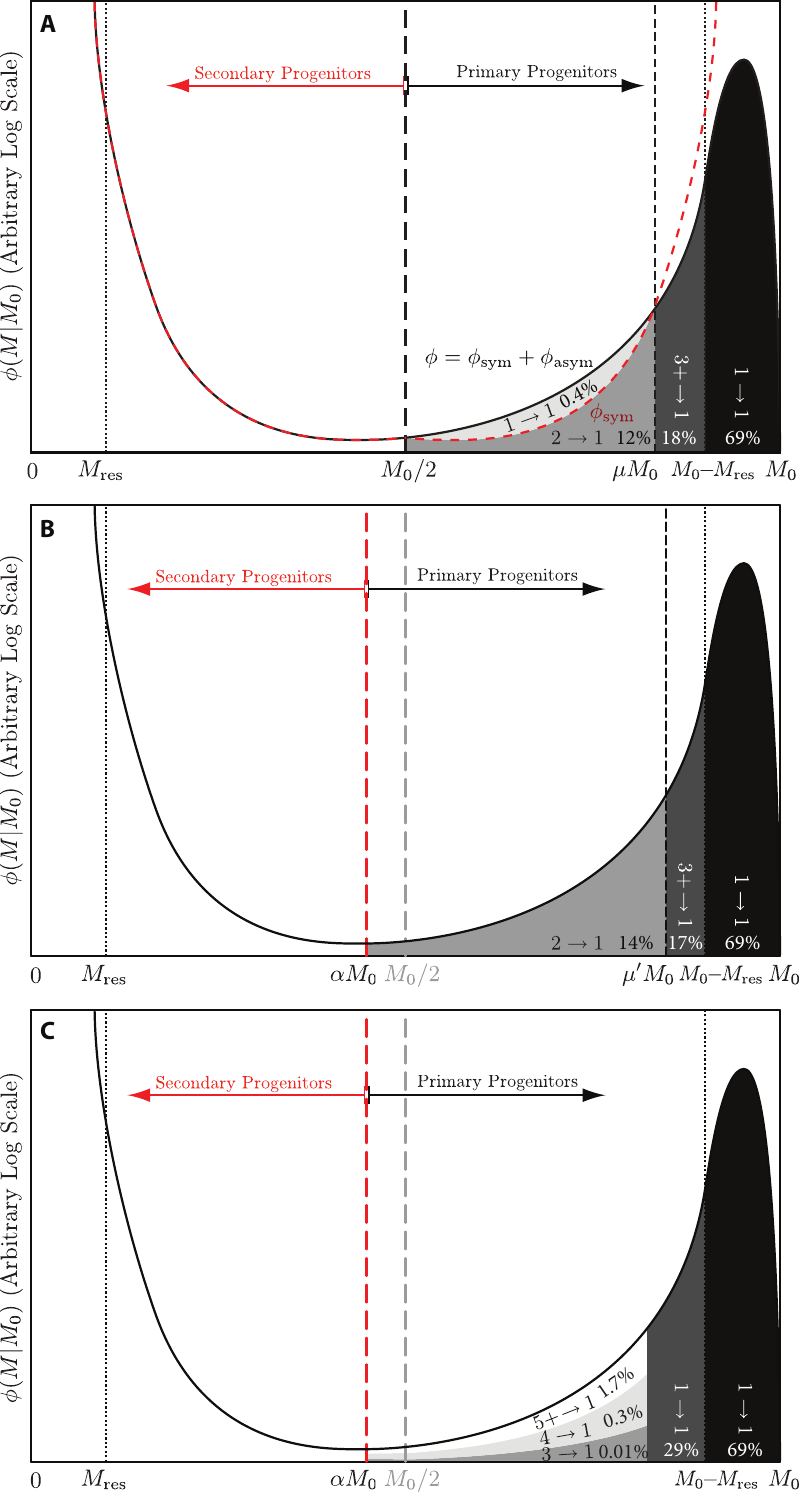} 
\caption{A schematic summary of how the three new algorithms proposed in
  this paper assign progenitors to descendants in a single time-step (see
  \S5).  The regions are shaded according to the progenitor multiplicity
  (marked by $N_p\rightarrow 1$) and the mass ranges.  See
  Table~\ref{newMethodsTableSummary} for a description of each shaded
  region and the fraction of descendants that belongs to each region. 
 The numbers quoted in this plot are from the ellipsoidal EPS model.
  The axes are in arbitrary units, though the horizontal axis is
  drawn to be symmetric about $M_0/2$ and the vertical axis is assumed to
  be logarithmic.  Important characteristic progenitor masses are labeled
  on the horizontal axis (see \S5.1.2 for discussion). The dashed line in
  panel $\textsf{A}$ plots $\Nsym$, the reflection of the left side of
  $\phi(M|M_0)$. 
}
\label{fig:methodsExplanation}
\end{figure}

Fig.~\ref{alpha_mu} shows $\alpha$ and $\mu$ as functions of the 
look-back time $\Delta z$ for three descendant halo masses 
($10^{12}M_{\odot}$, $10^{13}M_{\odot}$, $10^{14}M_{\odot}$) at redshift 
zero. According to the figure, $\alpha$ and $\mu$ have well defined constant 
values when $\Delta z$ is less than about $0.05$, a natural upper limit of 
time step-size for a Monte Carlo simulation to achieve convergence in 
both the spherical and ellipsoidal EPS models.

%%%%%%%%%%%%%%%%%%% Fig 6 %%%%%%%%%%%%%%

\subsubsection{Mass Bins} 
\label{massbin}

To help the reader reproduce our Monte Carlo algorithms, we discuss our
distribution of mass bins.

We divide the descendant mass range $10^6\leq M_0\leq10^{15}M_\odot$ into
$\sim100$ logarithmic descendant bins. Halos that fall into the same
descendant bin are assumed to have the same distribution of single-time-step
merger configurations that are computed using the central (in logarithmic
scale) value of the bin as the descendant mass. The progenitor masses in a
merger configuration are recorded in the form of ratios to the descendant
halo mass, instead of their absolute masses. This allows us to correct for
the (small) difference between the descendant halo in question and the
central mass of its bin.

For a given descendant mass $M_0$, its progenitor mass range $[\Mres, M_0]$
is divided into a certain number of mass bins to facilitate the process of
forming merger configurations. Interestingly, we note that simply dividing
the whole progenitor mass range into evenly spaced logarithmic bins is not
accurate, as discussed in \S\ref{KW93sec}. This is because the simplest
logarithmic binning assigns very few bins to the mass range of $[M_0/2,
M_0]$, which requires many mass bins to sample accurately the shape of the
sharply peaked (at around $M_0$) progenitor mass function for a small
time-step. To give the peaked region more fine structures, we find a simple
way: the mass range of $[\Mres, M_0/2]$ is divided into evenly spaced
logarithmic bins, and its reflection about the mid point $M_0/2$ determines
the binning on the right side of the mid point. Mathematically, it can be
stated as follows: The progenitor mass range $[\Mres,M_0]$ is divided into
$2N+1$ logarithmic mass bins. The $i^{th}$ ($i=0,1,2, ..., 2N$) bin spans
$[M^{i+1},M^i]$, where $M^i$ is defined as follows:
\[ M^{i}= \left \{ 
\begin{array}{ll}
	M_0 & \mbox{if $i=0$};\\
	\exp\left[ \ln \Mres+\Delta \times(2N+1-i)\right] & \mbox{if $i \geq N+1$};\\
	M_0-M^{2N+2-i} & \mbox{if $1\leq i\leq N$}. 
\end{array}
\right. \]
and $\Delta=(\ln (M_0/2)- \ln \Mres)/N$.

The average number of progenitors (per descendant halo) in the $i^{th}$ bin
is called $N^i$, which is equal to
$\int_{M^{i+1}}^{M^i}\phi(M|M_0)dM$. Note that $N^i$ is not an integer. For
$i\geq 1$, we choose the mean mass $\bar{M}^i$ of the $i^{th}$ bin to be
$\sqrt{M^iM^{i+1}}$. The progenitor mass function often changes rapidly
across the $0^{th}$ bin so we do not assign it a mean mass. Instead,
whenever a progenitor of the $0^{th}$ bin is needed, we generate a
probabilistic progenitor mass according to the progenitor mass function
inside this bin.

%%%%%%%%%%%%%%%%%%%%%%%%%%%% Table 2 %%%%%%%%%%%%%%%%%%%%%%%%%%%%%%
\begin{table}
	\centering 
	\begin{tabular}
		{r|ccc|ccc} & \multicolumn{3}{|c|}{Spherical EPS} & \multicolumn{3}{|c|}{Ellipsoidal EPS (z=0)} \\
		\hline $M_0$ ($M_\odot$) & $10^{12}$ & $10^{13}$ & $10^{14}$ & $10^{12}$ & $10^{13}$ & $10^{14}$ \\
		\hline $\alpha$ & 0.421 & 0.448 & 0.435 & 0.361 & 0.384 & 0.372 \\
		$\mu$ & 0.977 & 0.977 & 0.970 & 0.974 & 0.970 & 0.956 \\
		\hline 
	\end{tabular}
	\caption{Values of the progenitor mass scales $\alpha$ and $\mu$ discussed in \S\ref{mass_scales} for the spherical and ellipsoidal EPS models for three descendant masses ($10^{12}$, $10^{13}$, and $10^{14} M_{\odot}$) and $\dz=0.02$, where $\alpha M_0$ is defined such that $\int_{\alpha M_0}^{M_0} \phi(M|M_0)\,dM=1$ and $\mu M_0$ is defined such that $\phi(\mu M_0|M_0)=\phi(M_0-\mu M_0|M_0)$ with $\mu\neq 0.5$. } \label{boundaryStats} 
\end{table}

%%%%%%%%%%%%%%%%%%%%%%%%%%%% Table 3 %%%%%%%%%%%%%%%%%%%%%%%%%%%%%%
\begin{table*}
	
	\centering 
	\begin{tabular}
		{c|c|c|c|p{3.8in}|r}
		
		Method & $N_p$ & $\%$ Desc. & $\%$ Desc. & Description & Key \\
		& & (spher.) & (ellip.) & & \\
		
		\hline \hline A & $0\rightarrow1$ & 0.3\% & 0.4\% & Descendants with no progenitors because $\int_{M_0/2}^{M_0} \phi\, dM < 1$ & N/A \\
		\hline & $1\rightarrow1$ & 60\% & 69\% & $M_1 \in [M_0-\Mres, M_0]$: binary-turned-singles due to $M_2 < \Mres$ & 
		\includegraphics{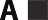} \\
		\hline & $1\rightarrow1$ & 0.8\% & 0.4\% & $M_1 \in [M_0/2, \mu M_0]$: $\Nsym < \phi$ results in unpaired primary progenitors: $\dM>\Mres$ & 
		\includegraphics{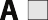} \\
		\hline & $2\rightarrow1$ & 21\% & 12\% & $M_1 \in [M_0/2, \mu M_0]$ and $M_0=M_1+M_2$: binary pairs generated from $\Nsym$ & 
		\includegraphics{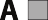} \\
		\hline & $3+\!\rightarrow1$ & 18\% & 18\% & $M_1 \in [\mu M_0, M_0-\Mres]$: $\Nsym > \phi $ results in excess secondary progenitors. $M_0<M_1+M_2+M_3+...$ & 
		\includegraphics{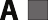} \\
		\hline \hline
		
		%---------------------------------------------------------------------------
		B & $1\rightarrow1$ & 60\% & 69\% & $M_1 \in [M_0-\Mres, M_0]$: binary-turned-singles due to $M_2 < \Mres$ & 
		\includegraphics{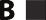} \\
		\hline & $2\rightarrow1$ & 20\% & 14\% & Binary paired progenitors generated by the iterative algorithm of \S\ref{methodB}: $M_0\geq M_1+M_2$ & 
		\includegraphics{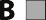} \\
		\hline & $3+\rightarrow1$ & 20\% & 17\% & $M_1 \in [\mu M_0, M_0-\Mres]$: identical to $3+\!\rightarrow1$ configuration in method A & 
		\includegraphics{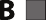} \\
		\hline \hline
		
		%---------------------------------------------------------------------------
		C & $1\rightarrow1$ & 60\% & 69\% & $M_1 \in [M_0-\Mres, M_0]$: binary-turned-singles due to $M_2 < \Mres$ & 
		\includegraphics{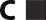} \\
		\hline & $1\rightarrow1$ & 35\% & 29\% & All secondary progenitors have already been assigned to smaller primary progenitors: these remaining primary progenitors have $\dM>\Mres$ & 
		\includegraphics{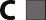} \\
		\hline & $3\rightarrow1$ & 0.1\% & 0.01\% & Merger configurations with 3 progenitors & 
		\includegraphics{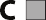} \\
		\hline & $4\rightarrow1$ & 2\% & 0.3\% & Merger configurations with 4 progenitors & 
		\includegraphics{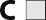} \\
		\hline & $5+\rightarrow1$ & 2.9\% & 1.7\% & Merger configurations with 5 or more 
                    progenitors & 
		\includegraphics{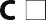} \\
		\hline \hline
		
		%---------------------------------------------------------------------------
		KW93 & $1\rightarrow1$ & 60\% & 69\% & $M_1 \in [M_0-\Mres, M_0]$: binary-turned-singles due to $M_2 < \Mres$ & N/A\\
		\hline & $1\rightarrow1$ & 15\% & 9\% & Merger configurations with a single progenitor with $M_1 < M_0-\Mres$ & N/A\\
		\hline & $2\rightarrow1$ & 11\% & 9\% & Merger configurations with 2 progenitors & N/A \\
		\hline & $3+\rightarrow1$ & 14\% & 13\% & Merger configurations with 3 or more progenitors & N/A \\
		\hline \hline
		
		%---------------------------------------------------------------------------
\end{tabular}

\caption{A summary of our three new Monte Carlo methods discussed in \S\ref{newmethods} 
  and the method of KW93. The percentages indicate the fractions of descendants 
  with $N_p$ progenitors in a given method, computed for $M_0=10^{13} M_\odot$ and 
  $\Mres=0.001M_0$ for a single time-step $\Delta z=0.02$ in both the spherical 
  and ellipsoidal EPS models. They are representative of the merger configuration 
  distributions for other descendant halo masses $M_0$.} 
\label{newMethodsTableSummary} 
\end{table*}

%%%%%%%%%%%%  S-EPS: new methods  \phi  %%%%%%%%%%%%%%%%%%%
\begin{figure*}
\centering 
\includegraphics{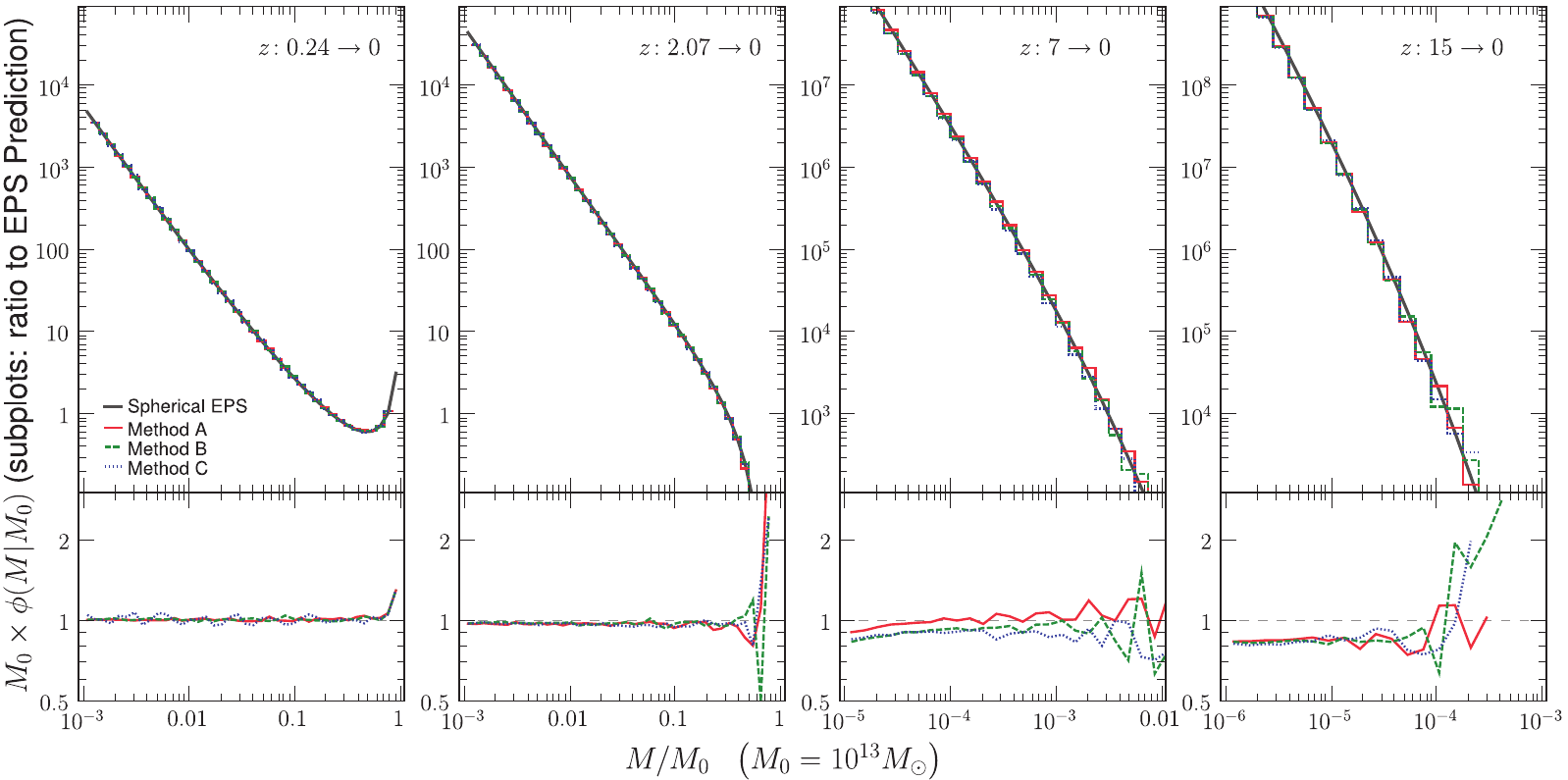} 
\caption{Comparison of the progenitor (or conditional) mass functions
  $\phi(M,z|M_0,z_0)$ generated using the three new Monte Carlo algorithms
  introduced in \S\ref{newmethods}: A (red solid), B (green dashed), and C
  (blue dotted), and the predictions of the spherical EPS model (black
  solid).  The four panels show four look-back redshifts ($z-z_0=0.24,
  2.07, 7, 15$) for a descendant halo of mass $10^{13}M_{\odot}$ at
  $z_0=0$.  For clarity, we plot in the sub-panel below each panel the
  ratios of the Monte Carlo result and the EPS prediction.  All three
  algorithms are seen to match very closely the spherical EPS prediction at
  all redshifts.  At $z=7$ and 15, the slight underestimates of the
  progenitor abundances at $M/M_0 \la 10^{-4}$ are primarily due to the
  fact that we trace a halo's progenitors only down to 0.001 of the halo
  mass in each small time-step in our Monte Carlo simulations.}
\label{PDF_ABC_e3}
\end{figure*}

%%%%%%%%%%%%  E-EPS: new methods  \phi  %%%%%%%%%%%%%%%%%%%
\begin{figure*}
\centering 
\includegraphics{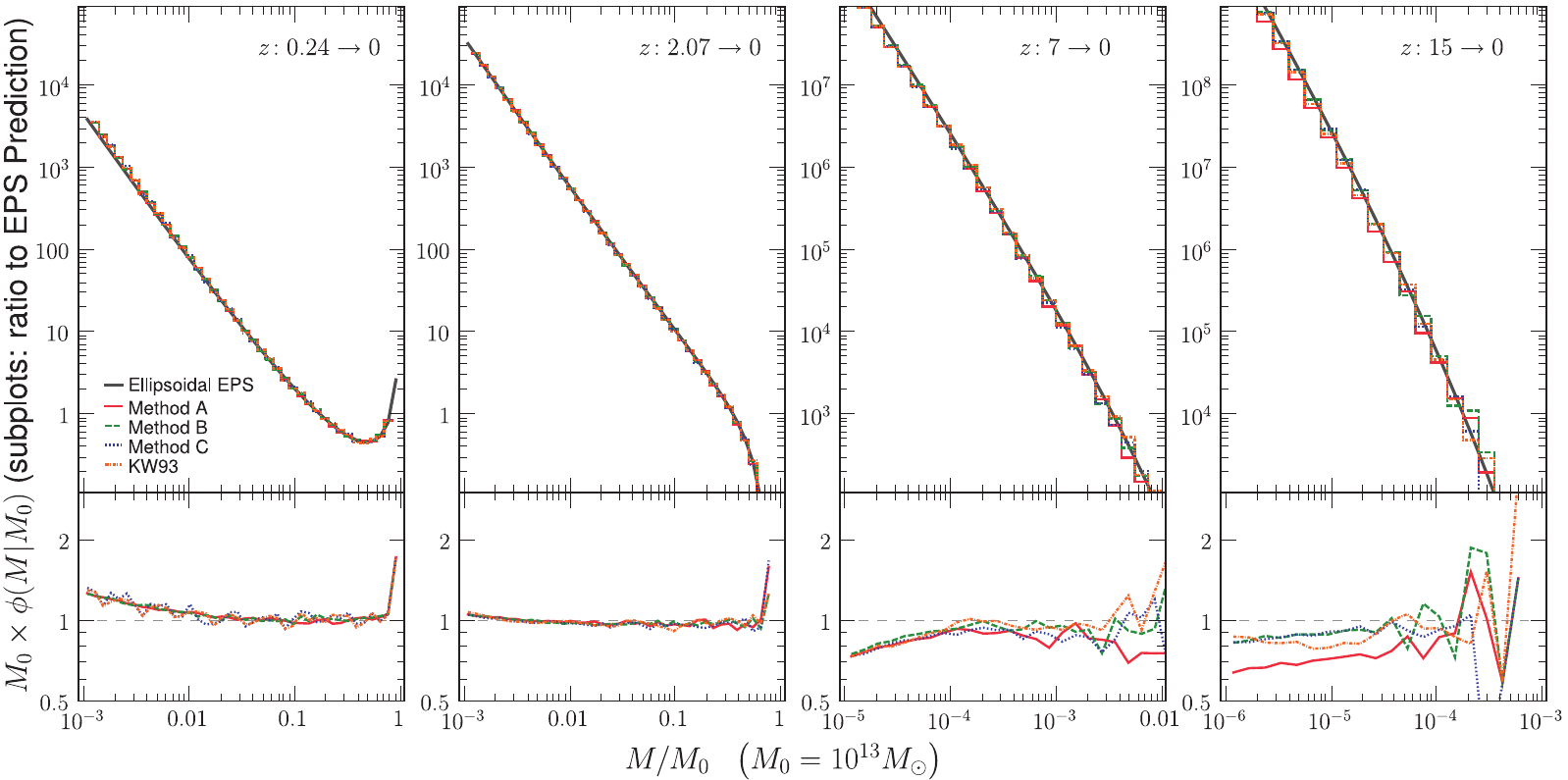}
\caption{Same as Fig.~\ref{PDF_ABC_e3} except both the Monte Carlo and
  analytic results are now generated from the {\it ellipsoidal} instead of
  the standard spherical EPS model. The Monte Carlo methods use
  eq.~(\ref{dndm_E}) as the progenitor mass function for each time step.
  The analytic results are calculated using the integral equation proposed
  by Zhang \& Hui (2006). The agreement is again excellent, indicating that
  our new Monte Carlo algorithms work well in reproducing the EPS
  progenitor mass function regardless if the EPS model is based on constant
  (i.e. spherical collapse) or moving barrier (ellipsoidal) random
  walks. For completeness, we include the results from the ellipsoidal
  version of the KW93 method (orange dash-dotted).  At z=0.24, the slight
  progenitor overabundance at the low mass end is due to the approximate
  nature of eq.~\ref{dndm_E}.  At z=7 and 15, the slight underestimates of
  the progenitor abundances are due to both the mass resolution issue as stated
  in the caption of Fig.~\ref{PDF_ABC_e3} and the barrier intersection
  problem of the ellipsoidal EPS model, which prevents us from tracing
  progenitors that are much smaller than the typical halo mass of the same
  redshift.}
\label{PDF_ABC_e3_E} 
\end{figure*}

\subsection{Method A} 
\label{methodA}

We first attempt to resolve the asymmetry problem in the EPS progenitor
mass function $\phi(M|M_0)$ by assuming that the primary progenitors in the
symmetric part $\Nsym$ in eq.~(\ref{asym}) are paired up with secondary
progenitors to form binary mergers such that $M_0=M_1+M_2$. This is done so
long as the smaller progenitor is above the mass resolution of the Monte
Carlo simulation, i.e. $M_2 \ge \Mres$ and $M_1 \le M_0 - \Mres$. If $M_2 <
\Mres$, then the second progenitor is discarded and $M_1$ is assumed to be
a single progenitor (the darkest grey region marked $1\rightarrow 1$ in
Fig.~\ref{fig:methodsExplanation} $\textsf{A}$). The remaining primary
progenitors in the asymmetric part $\Nasym$ are assumed not to pair up,
i.e. each descendant halo has a single progenitor (the lightest grey region
marked $1\rightarrow 1$ in Fig.~\ref{fig:methodsExplanation} $\textsf{A}$).

In practice, we generate the merger configurations of a descendant halo of
mass $M_0$ at each time step by repeating these two simple steps:

(i) Draw the primary progenitor mass $M_1$ from the mass range $[M_0/2,
M_0]$ of the progenitor mass function.

(ii) If $M_1 > M_0-\Mres$, no more progenitors are generated; if $M_1 \leq
M_0-\Mres$, the probability of having a second progenitor of mass
$M_2=M_0-M_1$ is set to
\begin{equation}
	r = \frac{\Nsym(M_1|M_0)}{\phi(M_1|M_0)} = \frac{\phi(M_0-M_1|M_0)}{\phi(M_1|M_0)}\,. 
\end{equation}
Then, drawing a random number $x$ between 0 and 1 allows us to determine
whether a secondary progenitor should be generated. If $x < r$, $M_2$ is
assigned as a secondary progenitor; otherwise $M_1$ is left as the sole
progenitor.

We point out two subtleties with this algorithm. First, $r$ is not always
$\le 1$. It is true that $r$ is below 1 for {\it most} of the relevant mass
range $M_1 \in [M_0/2, \mu M_0]$ (see Fig.~\ref{fig:methodsExplanation}
$\textsf{A}$ and Table~\ref{boundaryStats}) since the left side of
$\phi(M|M_0)$ is slightly lower than the right side. But when $M_1>\mu
M_0$, we find that $r\gsim 1$, implying that on average more than one
secondary progenitors should be generated to couple with the primary
progenitor $M_1$, and we must generate merger configurations with multiple
progenitors. To accommodate this feature, for each $M_1$ that satisfies $
M_1 \in [\mu M_0, M_0-\Mres]$, we generate\footnote{Here $\mathrm{int}(r)$
  is defined to be the largest integer $n$ that satisfies $n\leq r$.}
either $\mathrm{int}(r)$ or $\mathrm{int}(r)+1$ secondary progenitors of
mass $M_0-M_1$ according to whether a random number between 0 and 1 is
larger or smaller than $r-\mathrm{int}(r)$. Note that the resulting merger
configurations do not conserve mass exactly because the sum of the
progenitor masses is slightly larger than the descendant mass. Typically
most of these configurations only end up with 3 or 4 progenitors as $r
\lsim 2$ for $M_1 \lsim 0.999M_0$ and $\dz \lsim 0.02$.

The second subtlety with method A is that since the total number of
progenitors in the mass range of $[M_0/2, M_0]$ (which is equal to
$\int_{M_0/2}^{M_0}\phi(M|M_0)dM$) is always slightly smaller than one
(typically by 0.2\% to 0.4\% for $\dz =0.02$; recall from
Table~\ref{boundaryStats} that $\alpha M_0<M_0/2$), it is possible that we
sometimes cannot assign any progenitors to a given descendant halo. When
this happens, the descendant halo does not have any progenitor halos and is
a "$0\rightarrow1$ event".

For a thorough description of our algorithm A, we list below all the
possible merger configurations and their frequencies of occurrence for
descendant halos at $z=0$ over a single simulation time-step $\dz = 0.02$
and mass resolution $\Mres= 0.001 M_0$. This information is also summarized
in Table~\ref{newMethodsTableSummary} and
Fig.~\ref{fig:methodsExplanation}. In general, the relative frequencies of
different merger configurations are insensitive to the descendant mass
$M_0$ but do depend on the $\Delta z$ and $\Mres$ used in the Monte Carlo
simulation. For example, the fraction of $1\rightarrow1$ events increases
as $\Delta z$ decreases; and if $\Mres$ is chosen to be larger than
$(1-\mu) M_0 \sim 0.03 M_0$, then there are no $3\rightarrow1$ or
$4\rightarrow1$ mergers at each time-step and mass conservation is exactly
respected.

I. About 12\% in the ellipsoidal model (21\% for spherical) of descendant
halos have two progenitors each. These are binary pairs drawn from the
symmetric part of the progenitor mass function $\Nsym$, where $M_1 \in
[M_0/2, \mu M_0]$ and $M_0 = M_1 + M_2$ (Fig.~\ref{fig:methodsExplanation}
\includegraphics{A30.pdf}).

II. About 69\% (60\%) of descendant halos have only one progenitor each.
The majority ($\gsim 99\%$) of these descendants originally have binary
progenitors but the smaller progenitor is discarded since $M_2<\Mres$
(i.e. $M_1 \gsim M_0-\Mres$) (Fig.~\ref{fig:methodsExplanation}
\includegraphics{A80.pdf}). The rest ($\lsim 1\%$) of these
descendant have progenitors with $M_1 \in [M_0/2, \mu M_0]$ and originate
from the small asymmetric part $\Nasym$ of the progenitor mass function
where $r<1$ (Fig.~\ref{fig:methodsExplanation}
\includegraphics{A10.pdf}).

III. About 18\% of descendant halos have three or four progenitors each,
typically consisting of one massive progenitor and two or three very small
secondary progenitors ($\lsim (1-\mu) M_0 \sim 0.03M_0$). These have $M_1
\in [\mu M_0, M_0-\Mres]$ (Fig.~\ref{fig:methodsExplanation}
\includegraphics{A50.pdf}).

IV. About 0.4\% (0.3\%) of the descendants have no progenitors due to the
sharp cutoff of the primary progenitor mass at $M_0/2$ discussed above.

The red solid curves in Fig.~\ref{PDF_ABC_e3} compare the progenitor mass
functions from this Monte Carlo algorithm with the analytic
eq.~(\ref{dndm}) of the spherical EPS model. Fig.~\ref{PDF_ABC_e3_E} shows
the same thing except everything is for the ellipsoidal EPS model, where we
use eq.~(\ref{dndm_E}) to compute the progenitor mass function for each
small simulation time-step.  Both figures show excellent agreement ($< 10\%$
deviation) at $z_1-z_0=0.24$, 2.07, 7, and 15 for a descendant halo of mass
$10^{13}M_{\odot}$ at $z_0=0$. We have tested other descendant masses
($10^{12} M_\odot \la M_0 \la 10^{14} M_\odot$) and found equally good
agreement. This agreement also
provides numerical verification of the criterion introduced in
\S\ref{theorem}.

\subsection{Method B} 
\label{methodB}

Two features in method A may seem unnatural. First, as shown in
Table~\ref{newMethodsTableSummary} and discussed in the previous section, a
small fraction ($\sim 0.3\%$ to 0.4\%) of the descendant halos in method A
are not assigned any progenitors in one time-step because
$\int_{M_0/2}^{M_0}\phi(M|M_0)dM \approx 0.997$ (for $\Delta z = 0.02$ and
a large range of $M_0$) and is not exactly unity. It is important to note
that though these descendants are rare, one cannot remove them from method
A by modifying the normalization of $\phi(M|M_0)$ in the mass range of
$[M_0/2,M_0]$, as such a modification is amplified with iterations
and leads to a large error in $\phi(M|M_0)$ after many time-steps.

Second, due to the asymmetry in the EPS $\phi(M|M_0)$, we have assigned a
small fraction (0.4\% to 0.8\% for parameters used in
Table~\ref{newMethodsTableSummary}) of the descendant halos to
$1\rightarrow 1$ events. There is therefore a small chance that a
progenitor of mass comparable to half of the descendant mass does not have
any companions, corresponding to a large deficit between the mass of the
descendant halo and the total mass of its progenitors.

The first feature can be avoided by decreasing the lower limit of the mass
range from which the primary progenitor is drawn from $M_0/2$ to $\alpha
M_0$, where $\alpha$ is defined in eq.~(\ref{alpha}) and ranges from
$\alpha \approx 0.36$ to $0.45$ in Table~\ref{boundaryStats}.  The second
feature can be altered by distributing the secondary progenitors in a
slightly different way. These options motivate us to invent Method B with
the following set up:

{\bf 1.} We assume the primary progenitor mass lies in the mass range
$[\alpha M_0,M_0]$. This condition guarantees that every descendant halo
has a primary progenitor of mass $> \alpha M_0$ due to the definition of
$\alpha$.

{\bf 2.} We then assign secondary progenitors to primary progenitors from
the left side of $\alpha M_0$. For simplicity, whenever possible, we make
only binary configurations, each of which contains one primary and one
secondary progenitor. We start with the primary and secondary progenitor
bins that share the $\alpha M_0$ boundary (i.e. nearly equal-mass pairs)
and work our way outwards to the $M_1 \gg M_2$ pairs. This is a natural
decision as this way of pairing the primary and secondary masses minimizes
the difference between $M_0$ and $M_1+M_2$. Specifically, for a given $M_1$
bin, we determine its binary companion's mass $M_2$ from
\begin{equation}
	\int_{M_2}^{\alpha M_0} \phi(M|M_0)\,dM = \int_{\alpha M_0}^{M_1} \phi(M|M_0)\,dM \,, 
\label{Brelationship} 
\end{equation}
which guarantees that we always have an equal number of secondary
progenitors to pair up with the primary halos. Note that since $\alpha<0.5$
it is generally true that $M_0>M_1+M_2$.

{\bf 3.} One caveat with step 2 above is that this simple binary pairing
scheme works for a large range of masses but needs to be modified near the
end points when $M_1$ is close to $M_0$ and $M_2 \ll M_1$. This is because
the scheme starts out with nearly equal-mass pairs at $M_1\sim M_2 \sim
\alpha M_0$ and $M_1+M_2 < M_0$, and the asymmetric shape of the progenitor
mass function is such that the method produces pairs with increasing
$M_1+M_2$ as we move outward from $\alpha M_0$. The equality $M_1+M_2 =
M_0$ is reached when $M_1$ is slightly larger than $\mu M_0$ (typically at
$0.99M_0$), beyond which there are more secondary progenitors left to be
paired than the primary ones. We therefore stop the binary pairing when
$M_1+M_2=M_0$ is reached. From this point on, we instead use the same
multiple merger configurations as in method A. For simplicity in the
following few paragraphs, we denote this transitional $M_1$ as $\mu'M_0$.

In summary, methods A and B are closely related and are compared
side-by-side in Table~\ref{newMethodsTableSummary} and
Fig.~\ref{fig:methodsExplanation}. They have identical merger
configurations in the following regions:

I. The high-$M_1$ region $M_1 \in [M_0-\Mres, M_0]$, where 60\% to 70\% of
descendant halos belong. The secondary progenitor is below $\Mres$, so
$M_1$ is effectively the sole progenitor (i.e. $N_p=1$) for these
descendants

II. The region $M_1 \in [\mu' M_0, M_0-\Mres]$ ($\mu'$ replaced by $\mu$ in
method A), where 17\% to 20\% of descendant halos belong. These descendants all
each have 3 or more progenitors ($N_p=3+$).

Methods A and B differ in the following regions:

III. The binary pairing algorithm used in method B removes the sliver of
$1\rightarrow1$ configurations in the $M_1 \in [M_0/2, \mu M_0]$ region in
method A (\includegraphics{A10.pdf}) and redistributes the binary merger
configurations in this region (\includegraphics{A30.pdf}) to yield a robust
set of binary configurations between $\alpha M_0 \leq M_1 \leq \mu' M_0$
(\includegraphics{B30.pdf}). This affects $\sim 20$\% of the descendant
halos.

IV. Since the primary progenitor mass range extends down to $\alpha M_0$
instead of $M_0/2$, method B does not have any of the $0\rightarrow1$
configurations that are present in method A.

The green dashed curves in Figs.~\ref{PDF_ABC_e3} and \ref{PDF_ABC_e3_E}
compare the progenitor mass functions from method B with the analytic
predictions of the spherical and ellipsoidal EPS models, respectively. The
agreement is again excellent ($< 10\%$ deviation) at $z_1-z_0=0.24$, 2.07,
7, and 15 for a descendant halo of mass $10^{13}M_{\odot}$ at $z_0=0$.

Finally, we note that mass is not strictly conserved for the multiple
merger configurations generated in the $M_1 \in [\mu M_0, M_0-\Mres]$
region of method A and the $M_1 \in [\mu' M_0, M_0-\Mres]$ region of method
B (Fig.~\ref{fig:methodsExplanation}
\includegraphics{A50.pdf}, 
\includegraphics{B50.pdf}).  These configurations have more than
one companion of mass $M_0-M_1$, making the total mass of the progenitors
slightly above the descendant halo mass. This issue is due to the rapid
rise of the progenitor number as the secondary progenitor mass approaches
zero. In principle, the small progenitors ($\lsim (1-\mu) M_0$) that are
causing this problem can be re-distributed and combined, \eg, with
progenitors in some of the $1\rightarrow 1$ and $0\rightarrow 1$ merger
configurations in method A, or with some binary configurations of total
masses smaller than the descendant mass in method B, to form multiple
merger configurations that obey mass conservation (this, in fact, is what
happens in method C below, where mass conservation is strictly
respected). We have checked that this can be done successfully without
violating mass conservation down to very small $\Mres$ and find that in
practice, these modifications do not introduce significant changes to the
statistical properties of the halo merger histories. We have therefore
chosen to present the simpler version of each model. It is also worth
noting that in the EPS theory, mass conservation only has to be obeyed
statistically and is {\it not} required for individual merger
configurations.

\subsection{Method C (Multiple Mergers)} 
\label{methodC}

As shown in Table~\ref{newMethodsTableSummary}, methods A and B both
produce comparable number of descendants with binary ($N_p=2$) and multiple
($N_p=3+$) progenitors in a single time-step. The importance of multiple
merger configurations have been emphasized by a number of authors (e.g.,
\citealt{KW93,SK99,ND08b}). It is therefore interesting to explore the
relative importance of binary vs multiple mergers by relaxing the binary
assumption. Our method C is designed for this purpose. More specifically,
this method does not have any restrictions on the number of progenitors in
each merger configuration. We only require that the total progenitor mass
of every merger configuration be smaller than (or equal to) the descendant
halo mass.

We now describe method C:

{\bf 1.} To prevent the formation of $0\rightarrow1$ merger configurations
we mimic the setup of method B and choose to draw primary progenitors from
the mass range $M_1 \in [\alpha M_0,M_0]$. Thus methods B and C share the
same distribution of primary and secondary progenitor mass bins.

{\bf 2.} As with method B, we form merger configurations by assigning
secondary progenitors to progenitors in primary bins. Every primary bin
starts with one merger configuration: that which contains only the primary
progenitor itself, and has a probability $N_{\rm conf}$ equal to the number
of primary progenitors in the bin. The assignment of secondary progenitors
to primary bins is done in order of decreasing secondary progenitor
mass. For each secondary bin, we scan the primary bins in order of
increasing primary progenitor mass to find configurations with room to hold
\emph{at least} one secondary progenitor from the bin in question (recall
that we require the sum of progenitor masses to never exceed the descendant
mass).

{\bf 3.} When a valid configuration is found, we always assign the
\emph{maximal} number of secondary progenitors to that configuration. For
example, suppose we start to assign secondary progenitors from a bin with
central mass $M_2$ (say there are $N_2$ such progenitors in this bin), and
find a valid configuration of probability $N_{\rm conf}$ and total
progenitor mass $M_{tot}$. The maximum number $n_{max}$ of secondary
progenitors that can be added into each realization of this configuration
is equal to $\mathrm{int}[(M_0-M_{tot})/M_2]$. Therefore, we can maximally
assign $N_{max}=n_{max} \times N_{\rm conf}$ secondary progenitors to this
configuration.

I. If $N_{max} > N_2$, we break the configuration into two: one contains
the original set of progenitors, with a probability equal to
$(1-N_2/N_{max})\times N_{\rm conf}$; the other contains the original set
of progenitors plus $n_{max}$ secondary progenitors of mass $M_2$, with a
probability equal to $(N_2/N_{max})\times N_{\rm conf}$. In this case all
the secondary progenitors of the current secondary bin are assigned.

II. If $N_{max} \leq N_2$ we simply add the $n_{max}$ secondary progenitors
of mass $M_2$ to the configuration, and update the list of progenitors in
the configuration. $N_{\rm conf}$, the configuration's probability does not
change. The number of remaining secondary progenitors to be matched is now
$N_2-N_{max}$, and we continue our search across merger configurations (in
order of increasing primary progenitor mass) until all of them have been
assigned.

Once a secondary bin is fully assigned, we move on to the next secondary
bin (of a slightly smaller mass) and repeat the same assignment
procedure. As this process goes on all configurations are gradually filled
with secondary progenitors of smaller and smaller mass. For technical
convenience, the number of configurations in each primary bin and the
number of unique progenitor masses in each configuration are both limited
to be fewer than 6. In practice, we find that this setup allows us to
successfully assign all secondary progenitors in the mass range $[\Mres,
\alpha M_0]$, even when the mass resolution of each time step is as low as
$\Mres=0.001 M_0$.

In fact this dense packing of secondary progenitors into primary bin
configurations manages to distribute efficiently {\it all} secondary
progenitors in $[\Mres,\alpha M_0]$ in only a fraction of the available
primary progenitors. As seen in Fig.~\ref{fig:methodsExplanation}
$\textsf{C}$, only $2\%$ ($5\%$ for spherical) of the primary progenitors
(at the low mass end) are grouped with secondary progenitors and the
remaining $98\%$ ($95\%$) are $1\rightarrow1$ events. We note that even
though there are far more secondary progenitors than primary progenitors,
this is possible because many secondary progenitors have exceedingly small
masses and can be efficiently distributed into the mass reservoirs of
relatively few primary progenitors.

The execution of method C is as follows:

(i) Generate a primary progenitor $M_1$ from the mass range $[\alpha
M_0,M_0]$ of the EPS progenitor mass function. Determine which primary bin
contains $M_1$.

(ii) If $M_1 > M_0-\Mres$, no more progenitors are generated; if $M_1 \leq
M_0-\Mres$, a random number determines which merger configuration to choose
according to the probability distribution of all possible configurations
associated with the given primary bin. The progenitors of the chosen
configuration are then generated.

For a better understanding of method C, we show in
Table~\ref{newMethodsTableSummary} and discuss below all the possible
merger configurations and their frequencies of occurrence for descendant
halos (regardless of their masses) at $z=0$, assuming time-step $\Delta z =
0.02$ and mass resolution $\Mres= 0.001 M_0$:

I. About $98\%$ ($95\%$ for spherical) of the descendant halos have only
one progenitor each.

A) About $2/3$ of these descendants' progenitors are within the resolution
limit of the descendant mass (i.e. $M_1 \gsim M_0-\Mres$, see figure
\ref{fig:methodsExplanation}
\includegraphics{C80.pdf}).

B) The remaining $1/3$ of these descendant halos' progenitors have masses
below $M_0-\Mres$. As discussed above, these massive primary progenitors
are not assigned any secondary companions because all the available
secondary progenitors are maximally assigned to the less massive primary
bins. Note that this region extends to masses below $\mu M_0$ (
\includegraphics{C50.pdf}).

II. For the remaining primary progenitor bins, there are no configurations
having only two progenitors. All in all, $0.01\%$ ($0.1\%$ for spherical)
of all descendants have three progenitors (\includegraphics{C30.pdf});
$0.3\%$ ($2\%$) have four progenitors (\includegraphics{C20.pdf}); $1.7\%$
($2.9\%$) have five or more progenitors (\includegraphics{C10.pdf}). The
progenitor count for a given configuration can be rather large reaching
values of more than 100.

As in methods A and B, the values quoted above depend on $\Delta z$ and
$\Mres$. They also depend on the maximal number of configurations allowed
in each primary bin and the maximal number of unique progenitor masses
allowed in each configuration.

The blue dotted curves in Figs.~\ref{PDF_ABC_e3} and \ref{PDF_ABC_e3_E}
compare the progenitor mass functions from this Monte Carlo algorithm with
the analytic predictions of the spherical and ellipsoidal EPS models,
respectively. They again show excellent agreement ($< 10\%$ deviation) at
$z_1-z_0=0.24$, 2.07, 7, and 15 for a descendant halo of mass
$10^{13}M_{\odot}$ at $z_0=0$.

\section{Comparison of Higher-Moment Statistics in Algorithms A, B, C, and
  KW93}
\label{other_stat}

We have designed Monte Carlo algorithms A, B, and C for constructing merger
trees that can accurately reproduce the EPS prediction for the progenitor
mass function $\phi(M|M_0)$ across each individual time-step. According to
the discussion in \S\ref{theorem}, these methods should then accurately
generate the progenitor mass function at {\it any} look-back time in any
number of time-steps. Figs.~\ref{PDF_ABC_e3} and \ref{PDF_ABC_e3_E} show
that this is indeed the case for both the spherical and ellipsoidal EPS
models. Including KW93, there are now four methods that are completely
consistent with the EPS $\phi(M|M_0)$. The results of the ellipsoidal version 
of KW93 have been shown in Fig.~\ref{PDF_ABC_e3_E} as well.

Despite this agreement, we recall that the progenitor mass function is only
one of many statistical properties of a halo merger tree. Even though all
four algorithms are degenerate in $\phi(M|M_0)$, they are likely to (and
should) differ in their predictions for other statistical quantities. Here
we investigate two such quantities as an illustration: (i)
$\phi^{(N_p)}(M|M_0)$, the progenitor mass function for the subset of
descendant halos that have $N_p$ progenitors. The sum of
$\phi^{(N_p)}(M|M_0)$ over all $N_p$ is equal to $\phi(M|M_0)$. (ii)
$\phi^{(i_{th})}(M|M_0)$, the distribution of the $i_{th}$ most massive
progenitor of each descendant halo. Again, the sum of
$\phi^{(i_{th})}(M|M_0)$ over all $i$ is equal to $\phi(M|M_0)$.  These two
statistics are two obvious ways of decomposing the total $\phi(M|M_0)$ into
individual moments: $\phi^{(N_p)}$ separates flourishing trees from
quiescent trees, while $\phi^{(i_{th})}$ compares the individual
distributions of the primary, secondary and more minor progenitors, which
are relevant for modeling galaxy formation through mergers (see also
\citealt{PCH08}).  Other statistics such as the distributions of halo
formation time and last major merger time (e.g., \citealt{PCH08,C08,MS07})
and the factorial moments of the partition function (e.g.,
\citealt{SP97,SL99}) are also useful.  Some of these will be examined in
our next paper.

\begin{figure}
\centering 
\includegraphics{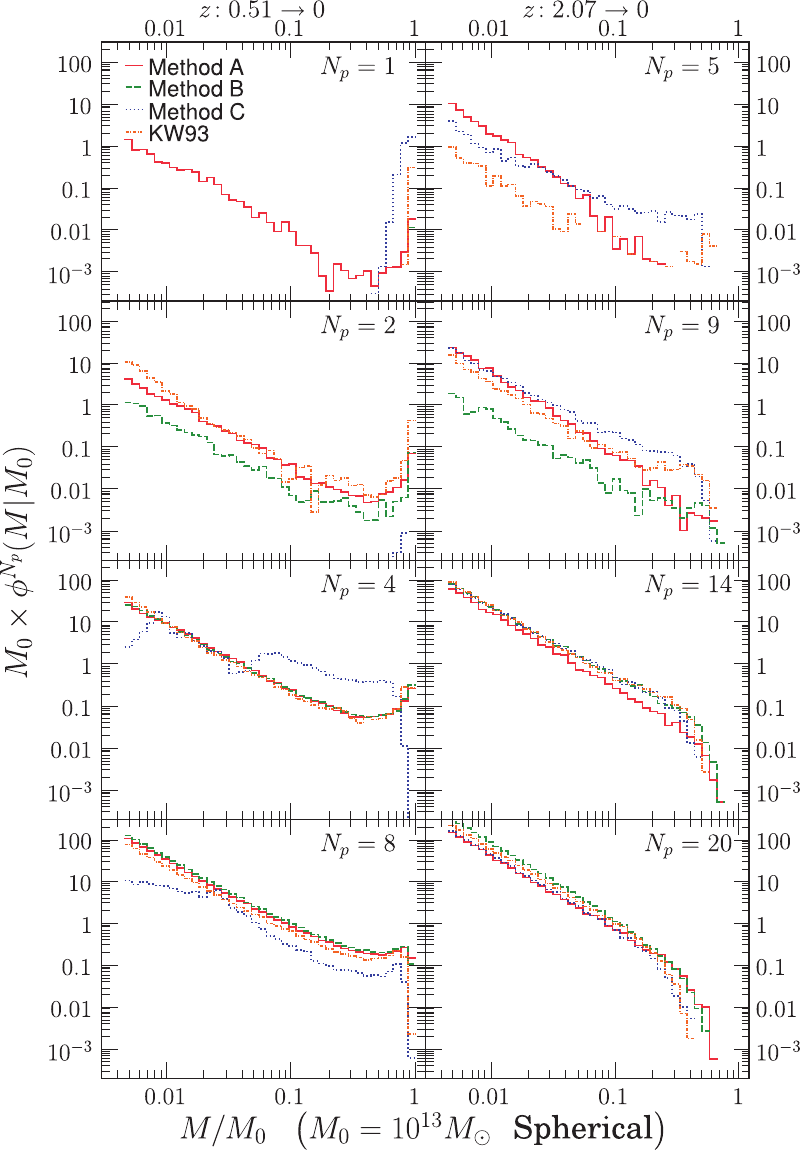} 
\caption{Predictions of algorithms A (red solid), B (green dashed), C (blue
  dotted), and KW93 (orange dash-dotted) for $\phi^{(N_p)}(M,z|M_0,z_0)$,
  the mass function of progenitors for descendant halos that have a total
  of $N_p$ progenitors.  Two look-back redshifts are shown: $z-z_0=0.51$
  (left) and 2.07 (right).  For each redshift, four representative values
  of $N_p$ are shown (from top down).  The simulations are for the
  spherical EPS model and assume a descendant halo mass of
  $10^{13}M_{\odot}$ at $z_0 =0$ and mass resolution of $\Mres=4\times
  10^{10}M_{\odot}$.  }
\label{PDFN_ABC_e3_S}
\end{figure}

\begin{figure}
\centering 
\includegraphics{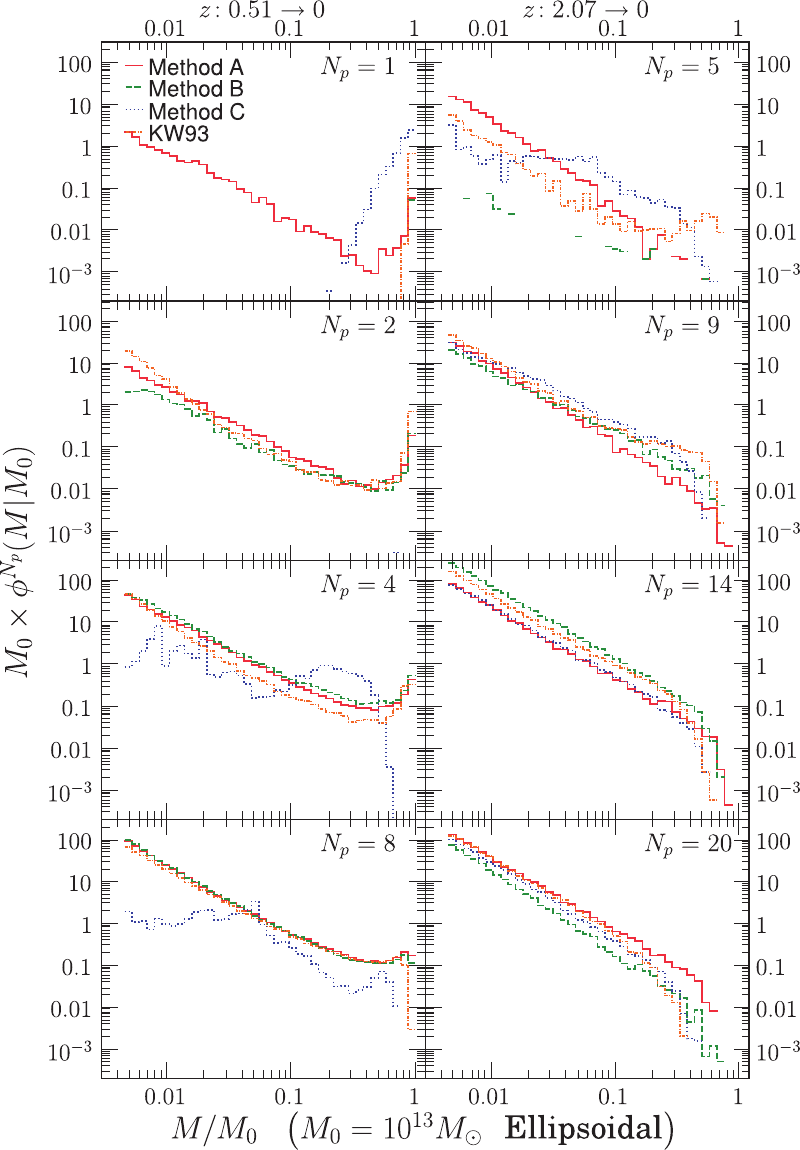}
\caption{Same as Fig.~\ref{PDFN_ABC_e3_S} except the Monte Carlo results
  are generated from the ellipsoidal instead of the standard
  spherical EPS model. } \label{PDFN_ABC_e3_E}
\end{figure}

\begin{figure}
\centering 
\includegraphics{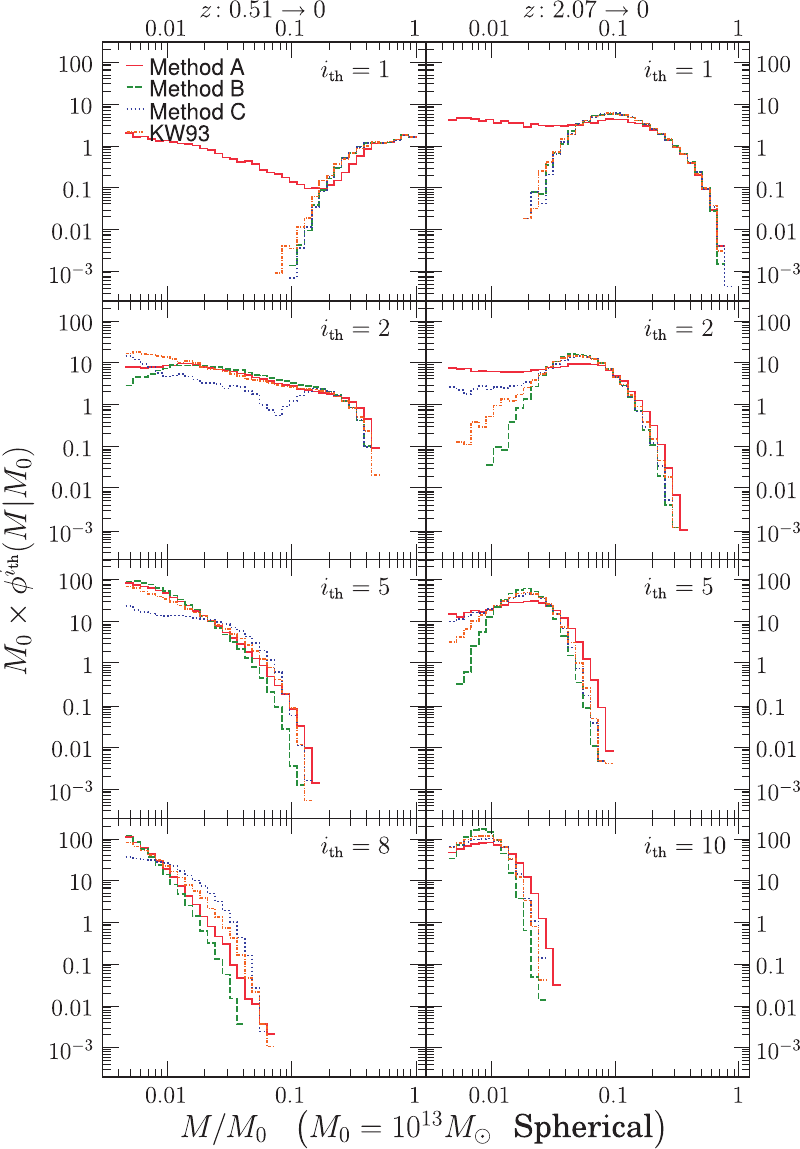} 
\caption{Same as Fig.~\ref{PDFN_ABC_e3_S} except for a different progenitor
  statistic: $\phi^{(i_{\rm th})}(M,z|M_0,z_0)$, the mass function of the
  $i_{th}$ most massive progenitor. 
} 
\label{PDFNth_ABC_e3_S}
\end{figure}

\begin{figure}
\centering 
\includegraphics{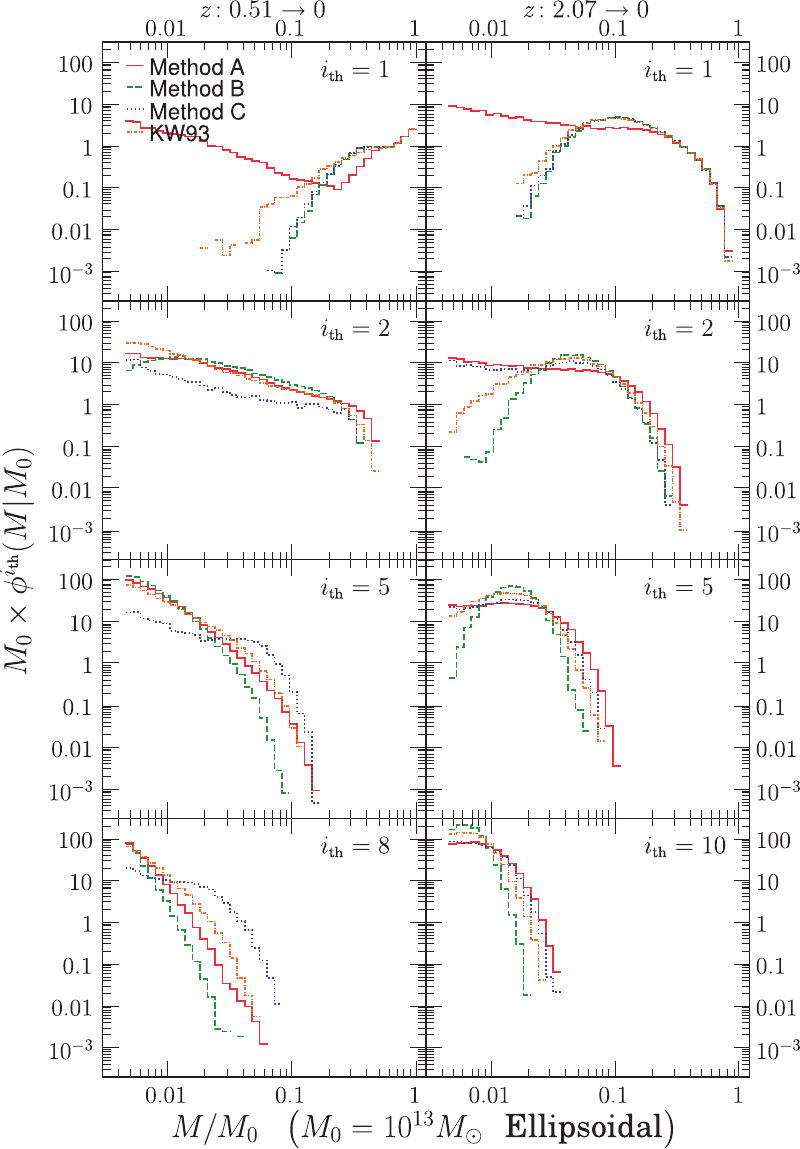} 
\caption{Same as Fig.~\ref{PDFNth_ABC_e3_S} except the Monte Carlo results
  are generated from the ellipsoidal instead of the spherical EPS
  model.}
\label{PDFNth_ABC_e3_E}
\end{figure}

To compute these moments, we set the descendant halo at redshift zero to be
$10^{13}M_{\odot}$, and the mass resolution to be $4\times
10^{10}M_{\odot}$. The results are plotted at two look-back times
($z_1-z_0= 0.51, 2.07$) in Figs.~\ref{PDFN_ABC_e3_S}-\ref{PDFNth_ABC_e3_E},
where Figs.~\ref{PDFN_ABC_e3_S} and \ref{PDFN_ABC_e3_E} show
$\phi^{(N_p)}(M|M_0)$ for the spherical and ellipsoidal EPS models,
respectively, while Figs.~\ref{PDFNth_ABC_e3_S} and \ref{PDFNth_ABC_e3_E}
show $\phi^{(i_{th})}(M|M_0)$.  In each figure, results from our three
methods (red solid for A, green dashed for B, blue dotted for C) and from
our implementation of KW93 (orange dash-dotted) are shown for comparison.
These figures clearly indicate that methods A, B, C, and KW93 generate
distinct predictions for these specific moments of the progenitor mass
distribution.  Some of the notable differences are:

{\bf 1.} Method C produces a much lower amplitude for the $N_p=2$ and 3
moments than methods A and B. This is because C is designed to be a
multiple-merger algorithm that effectively does not generate {\it any} binary
configuration in one individual time-step (note the absence of the $N_p=2$
entry for method C in Table~\ref{newMethodsTableSummary}). This feature
can been seen by the absence of blue short-dashed curves in the $N_p=2$ and
3 panels in Figs.~\ref{PDFN_ABC_e3_S} and \ref{PDFN_ABC_e3_E}, i.e., there
are almost no descendant halos having only two or three progenitors in
method C at $z=0.51$. By contrast, methods A and B have a wealth of
descendants with binary progenitors at these redshifts.

{\bf 2.} The removal of the binary assumption in method C leads to many
features in the moments of the progenitor distributions. By contrast, the
predictions from A and B are mostly power-laws, or at least smooth
functions, in the progenitor mass. This difference is due to the fact that
the merger configurations in the binary methods are much more regulated
than those in the non-binary method: a binary configuration contains only
two progenitors, the total mass of which is always quite close (if not
equal) to the descendant mass, whereas the distribution of progenitor
masses in a multiple configuration can have various forms, which can easily
affect, \eg, the ranking of the progenitor masses and the number of
progenitors. It is interesting to note that the predictions of KW93 are
fairly smooth functions in spite of the fact that it does not assume
binary. This is likely because the way progenitors are assigned in KW93
effectively suppresses the probability of mergers involving multiple
progenitors.

{\bf 3.} The differences between method A and B are more subtle because
they are both mostly binary methods. The main feature that distinguishes A
from B is in the distribution of the most massive progenitor
(i.e. $i_{th}=1_{}$) shown in the first columns of
Figs.~\ref{PDFNth_ABC_e3_S} and \ref{PDFNth_ABC_e3_E}. At the high mass
end, method B has a slightly broader shape for the primary progenitor mass
than method A. This is expected, because it is the case across every time
step by construction (the primary bins in method B extend down to $\alpha
M_0$ as opposed to $M_0/2$ for method A). At the low mass end, however,
there is a long tail in the distribution of primary progenitor masses in
method A, which is not present in other methods. This tail is caused by the
fact that in method A, there is a small chance ($\sim 0.3\%$) at every
time-step that a primary progenitor completely disappears, transferring the
rank of ``primary'' to one of the much smaller secondary progenitors. Over
several time-steps this rare occurrence affects more and more branches of the
merger tree and can significantly modify the primary progenitor statistics.

In summary, we have constructed three Monte Carlo algorithms that can all
reproduce closely the progenitor mass function of the EPS model (both
spherical and ellipsoidal). The methods, however, produce significantly
different higher moments of the progenitor distributions. They are also
very different from KW93. Either a theoretical model more complete than the
EPS or direct $N$-body results will be needed to determine which, if any,
of the thus-far successful algorithms is the winner. We will turn to this
subject in the next paper (Zhang, Fakhouri \& Ma, in preparation).

%%%%%%%%%%%%  S-EPS  N^(N_p)  %%%%%%%%%%%%%%%%%%%

%%%%%%%%%%%%  E-EPS  N^(N_p)  %%%%%%%%%%%%%%%%%%%

%%%%%%%%%%%%  S-EPS  N^(i_th)  %%%%%%%%%%%%%%%%%%%

%%%%%%%%%%%%  E-EPS  N^(i_th)  %%%%%%%%%%%%%%%%%%%

\section{Conclusions and Discussion} 
\label{summary}

Monte Carlo algorithms based on the spherical EPS model have been an
essential tool for many studies of galaxy and structure formation. These
algorithms allow one to generate realizations of actual halo merger
histories starting from a limited set of statistical information about dark
matter halo properties provided by the EPS model. Since the EPS model does
not uniquely determine many statistical quantities of halo mergers beyond
the progenitor mass function, there is considerable freedom in how to
combine progenitors to form descendant halos in each time step in a Monte
Carlo algorithm.

The emphasis of this paper is on elucidating and quantifying the ability of
a Monte Carlo algorithm to construct merger trees that match the analytic
progenitor mass function of the EPS model (both the spherical and
ellipsoidal versions). Four main conclusions can be drawn:

{\bf 1.} We have shown rigorously that to match the EPS progenitor mass
function accurately at any look-back time, it is necessary and {\it
  sufficient} for a Monte Carlo algorithm to reproduce the exact progenitor
mass function at each time step.

{\bf 2.} We have reviewed and compared the four most frequently used Monte
Carlo algorithms based on the spherical EPS model in the literature:
\citealt{LC93}, \citealt{KW93}, \citealt{SK99}, and \citealt{C00}. As seen
in Figs.~\ref{PDF_4EPS_e2}-\ref{PDF_4EPS_e4}, all but KW93 only
approximately reproduce the spherical EPS progenitor mass function at each time
step, resulting in large deviations from the spherical EPS predictions after the
accumulation of small errors over many time steps.

Their problems (see Table~\ref{methodsTableSummary} for details) can be
summarized as: (i) SK99 generally over-estimates the abundances of small
progenitors by about a factor of two; (ii) LC93 over-produces progenitors
by a factor of a few when the look-back time is large ($\Delta z\gg 1$);
(iii) C00 under-predicts the progenitor abundance at the high mass end when
the look-back time is large. The origin of these discrepancies frequently
comes from the incompatibility between the binary merger assumption used in
the Monte Carlo algorithm (e.g. LC93, C00) and the asymmetric progenitor
mass function of the EPS model.

{\bf 3.} We have designed three new Monte Carlo algorithms that all
reproduce closely the EPS progenitor mass function over a broad range of
redshift ($z_1-z_0$ up to at least 15) and halo mass.  Our methods A and B
assign binary pairs to the symmetric part of $\phi(M|M_0)$ and non-binaries
to the asymmetric part; the two differ in the mass ranges for the most
massive progenitors.  Our method C, on the other hand, completely relaxes
the binary merger assumption.  The algorithms are tested for both the
spherical and ellipsoidal EPS models and the results are shown in
Figs.\ref{PDF_ABC_e3} and \ref{PDF_ABC_e3_E}.  We see that all three methods
perform equally well at reproducing the respective progenitor mass function
at higher redshifts, regardless of whether the spherical progenitor mass
function eq.~(\ref{dndm}) or ellipsoidal progenitor mass function
eq.~(\ref{dndm_E}) is used as input.

{\bf 4.} As emphasized throughout the paper, the EPS model only provides a
partial statistical description of dark matter halo properties; it does not
tell us explicitly how to group progenitors into descendants in a Monte
Carlo realization. Therefore, there are different ways to combine
progenitors into descendant halos in consistent Monte Carlo algorithms.

We have used our three new algorithms to illustrate this exact
point. Despite their success in generating merger trees that accurately
reproduce the EPS progenitor mass function,
Figs.~\ref{PDFN_ABC_e3_S}-\ref{PDFNth_ABC_e3_E} show that the three
algorithms make significantly different predictions for quantities such as
the distribution of the most (or the $2_{nd}$ or $3_{rd}$ most) massive
progenitor masses, and the mass function of progenitors in descendant halos
with $N_p$ ($=1,2,3 ...$) progenitors. A theory more complete than EPS
would be needed to predict these higher-order merger statistics and break
the degeneracies in the progenitor mass function. Alternatively,
comparisons with $N$-body simulations should determine which, if any, of
the three new algorithms is viable. We view the EPS models (spherical or
ellipsoidal), Monte Carlo algorithms, and $N$-body simulations as three
major components in the general study of the formation, growth, and
clustering of dark matter halos. In this paper we have focused on the first
two areas, comparing various Monte Carlo algorithms for generating halo
merger trees and quantifying their abilities to consistently match the
analytical EPS progenitor mass functions over a broad range of mass and
redshift. In our next paper (Zhang, Fakhouri, Ma 2008b), we will turn to
comparisons with the Millennium simulation.

Several recent papers have investigated other Monte Carlo methods (see,
e.g., \citealt{PCH08, ND08a, MS07, ND08b}.  Although a complete review of
these methods is beyond the scope of this paper, it is worth pointing out
some of their features. The method of \cite{MS07} is essentially equivalent
to LC93 but is based on the ellipsoidal collapse model\footnote{They 
use a square-root approximation for the moving barrier form, which avoids
 the barrier crossing problem at the low mass end.} and is
  discretized in mass instead of redshift.  The two progenitor masses for
each time step are assigned using computer-generated random walks and
moving barriers. Since the asymmetry problem of the progenitor mass
function is also present in the ellipsoidal model, this method does not
accurately reproduce the theory-predicted progenitor mass function at each
time step. Such a discrepancy is amplified with increasing redshift and is
indeed shown in Fig.~5, 6, and 7 of \cite{MS07}.

\cite{ND08b} have proposed a method that exactly reproduces the progenitor
mass function of the spherical EPS model at each time step. This feature
alone guarantees it to be consistent with EPS at any look-back time
according to our discussion in \S3.1. However, since the method requires
solving several differential equations with nontrivial boundary conditions
for the progenitor masses, it is technically harder to implement it.

Finally, the methods described in \cite{PCH08} and \cite{ND08a} are
proposed to mimic N-body results. They are based on fitting to N-body data
rather than the EPS theory. It will be interesting to compare the
predictions for the various merger statistics discussed in this paper from
these methods with those from our ellipsoidal EPS-based methods and from
N-body simulations. This will be done in the next paper.

\section*{Acknowledgments}

We thank Michael Boylan-Kolchin, Lam Hui, Ravi Sheth, and Simon White for
discussions, and Eyal Neistein and Jorge Moreno for useful comments on an
earlier version of this paper. JZ is supported by NASA and by the TAC
Fellowship of UC Berkeley.  OF and CPM are supported in part by NSF grant
AST 0407351.

\bibliographystyle{mn2e}

\label{lastpage}

\end{document}